\pdfoutput=1
\newcommand*{\ATLASLATEXPATH}{}
\documentclass[atlasdraft=true, texlive=2016, USenglish, cernpreprint]{\ATLASLATEXPATH atlasdoc}

\usepackage{\ATLASLATEXPATH atlaspackage}
\usepackage{\ATLASLATEXPATH atlasbiblatex}

\usepackage{\ATLASLATEXPATH atlasphysics}
\usepackage{units}

\usepackage{paper2017-defs}

\AtlasTitle{Search for photonic signatures of gauge-mediated supersymmetry in \unit[13]{TeV} $pp$ collisions with the ATLAS detector}

\AtlasAbstract{%
A search is presented for photonic signatures, motivated
by generalized models of gauge-mediated supersymmetry breaking. This
search makes use of proton--proton
collision data at $\sqrt{s} = \unit[13]{TeV}$ corresponding to an integrated luminosity of \integLumi recorded by the ATLAS detector at the LHC, and it explores
models dominated by both strong and electroweak production of supersymmetric
partner states. Experimental signatures incorporating an isolated photon and
significant missing transverse momentum are explored. These signatures include events with
an additional photon or additional jet activity
not associated with any specific underlying quark flavor.
No significant excess of events is observed above the Standard Model prediction,
and 95\%~confidence-level upper limits of between 0.083~fb and 0.32~fb are set
on the visible cross section of contributions from physics beyond the Standard Model. These
results are interpreted in terms of lower limits on the masses of gluinos,
squarks, and gauginos in the context of generalized models of gauge-mediated
supersymmetry, which reach as high as \unit[2.3]{TeV} for strongly produced
and \unit[1.3]{TeV} for weakly produced supersymmetric partner pairs.
}

\AtlasRefCode{SUSY-2016-27}

\PreprintIdNumber{CERN-EP-2017-323}

\AtlasJournalRef{Phys. Rev. D 97 (2018) 092006}
\AtlasDOI{10.1103/PhysRevD.97.092006}

\AtlasCoverSupportingNote{Diphoton:}{https://cds.cern.ch/record/2233741}
\AtlasCoverSupportingNote{Photon+jets:}{https://cds.cern.ch/record/2233238}
\AtlasCoverEgroupEditors{atlas-SUSY-2016-27-editors@cern.ch}

\AtlasCoverEgroupEdBoard{atlas-SUSY-2016-27-editorial-board@cern.ch}

\hypersetup{pdftitle={ATLAS document},pdfauthor={The ATLAS Collaboration}}

\begin{document}

\maketitle

\tableofcontents

\section{Introduction}
\label{sec:intro}

This paper reports on a search for two complementary classes of events
containing energetic isolated photons and large missing transverse momentum
(with magnitude denoted \MET). The search is performed with
proton--proton ($pp$) collision data at a center-of-mass energy $\sqrt{s}=13$~TeV
corresponding to an integrated luminosity of \integLumi
recorded with the ATLAS detector at the Large Hadron Collider (LHC)
in 2015 and 2016. For the first of the two classes, two isolated energetic
photons are required (``diphoton'' events), while for the second class
only a single isolated photon is required, in combination with multiple
hadronic jets (``photon+jets'' events). 

The results of searches for these two classes of events are
interpreted in the context of several general models
of gauge-mediated supersymmetry breaking
(GGM)~\cite{Cheung:2007es,Meade:2008wd}.
These models include both the production of supersymmetric partners of
strongly coupled Standard Model (SM) particles
and the production of partners of SM particles possessing only electroweak charge.
In all models of GGM, the lightest supersymmetric particle (LSP)
is the gravitino \gravitino (the partner of the hypothetical quantum of the
gravitational field), with a mass significantly less than \unit[1]{GeV}.
In the GGM models considered here, the decay of the supersymmetric states produced in
LHC collisions would proceed through the next-to-lightest supersymmetric particle
(NLSP), which would then decay to the \gravitino LSP and one or more SM particles.
Each of the two event classes corresponds to a specific choice of NLSP, 
each of which in turn has a high probability of decay into $\gamma$ + \gravitino.
In all models considered, all supersymmetric states with the exception of the \gravitino
are short lived, leading to prompt production of SM particles that are observed in the
ATLAS detector.
The result based on the diphoton signature extends and supplants an ATLAS search~\cite{SUSY-2016-04}
performed with an integrated luminosity of \unit[3.2]{\ifb} of $pp$ collision data taken
at a center-of-mass energy of $\sqrt{s} =$ \unit[13]{TeV},
and complements searches~\cite{CMS-SUS-16-047,CMS-SUS-16-046} performed by the CMS Collaboration 
making use of \unit[35.9]{\ifb} of $\sqrt{s} =$ \unit[13]{TeV} $pp$ collision data.
The result based on the photon+jets signature extends and
supplants an ATLAS search~\cite{SUSY-2014-01} performed with an integrated luminosity of \unit[20.3]{fb$^{-1}$} 
of \unit[8]{TeV} $pp$ collision data.

The paper is organized as follows. More details of the theoretical background are 
provided in Section~\ref{sec:susy}. Section~\ref{sec:detector} presents the salient features of the ATLAS detector.
Section~\ref{sec:mc} provides details of the Monte Carlo simulations used in the analysis for background and signal processes.
Section~\ref{sec:objects} discusses the reconstruction and identification of photons, leptons, jets, and whole-event
observables relevant to the event selection, while Section~\ref{sec:selection} describes the event selection itself. 
The estimation of background contributions and signal efficiency,
and the study of systematic uncertainties are discussed in Sections~\ref{sec:background} and~\ref{sec:sig_eff}. 
The results are presented in
Section~\ref{sec:result} and are interpreted in terms of limits on various GGM models.
Finally, Section~\ref{sec:conclusion} is devoted to the conclusions.

\section{Gauge-mediated supersymmetry phenomenology}
\label{sec:susy}
\newcommand{\StopDecFootnote}{For the case of left-handed top squark (stop) production 
when $m_{\mathrm{stop}} < m_{\neutralino} + m_{\mathrm{top}}$, the stop
decay proceeds through an effective neutral current interaction to
a charm or up quark accompanied by the binolike \neutralino.}  

Supersymmetry (SUSY) \cite{Golfand:1971iw,Volkov:1973ix,Wess:1974tw,Wess:1974jb,Ferrara:1974pu,Salam:1974ig,Martin:1997ns,Martin:2000ns}
introduces a symmetry between fermions and bosons, resulting in a SUSY
partner (sparticle) for each SM particle with identical quantum numbers except a difference
by half a unit of spin. As none
of these sparticles have been observed, SUSY must be a broken symmetry
if realized in nature.  Assuming $R$-parity
conservation~\cite{Fayet:1976et,Fayet:1977yc,Farrar:1978xj,Fayet:1979sa,Dimopoulos:1981zb},
sparticles are produced in pairs.  These then decay through cascades involving
other sparticles until the stable, weakly interacting LSP is produced, leading
to a final state with significant \MET.

This paper considers experimental signatures associated with models
inspired by gauge-mediated SUSY 
breaking~\cite{Dine:1981gu,AlvarezGaume:1981wy,Nappi:1982hm,Dine:1993yw, Dine:1994vc,Dine:1995ag}.
These signatures are largely determined by the nature of the NLSP;
in GGM models, the NLSP is often formed from an admixture of any of the SUSY partners
of the electroweak gauge and Higgs bosons.
In this study, two cases are considered for the
composition of the NLSP, both of which would
produce photonic signatures in the ATLAS detector. In the first case, the NLSP is assumed to be
purely binolike [the SUSY partner of the SM U(1) gauge boson], while in the
second case, the NLSP is assumed to be an admixture
of bino and neutral higgsino states.
In this paper, the neutral NLSP is denoted $\neutralino$ irrespective
of its composition. 

Where not explicitly constrained by the assumptions of
the specific GGM models under study, the masses and properties of 
SUSY partner states are controlled by several underlying parameters. 
These include the U(1), SU(2) and SU(3) gauge partner mass
parameters ($M_1$, $M_2$ and $M_3$, respectively), the 
higgsino mass parameter $\mu$, the gravitino mass, 
and the ratio
$\tan \beta$ of the two SUSY Higgs-doublet vacuum expectation values.
A value of 1.5 is chosen for the latter;
for all GGM models considered, the
phenomenology relevant to this search is only
weakly dependent on the value of $\tan \beta$.

If the NLSP is binolike,
the final decay in each of the two cascades in a GGM SUSY event is predominantly
$\neutralino\to\gamma+\gravitino$, leading
to final states with two photons and missing transverse momentum.
If the NLSP is a mixture of the bino
and higgsino, the higgsino mass parameter $\mu$ is chosen
to be positive, leading to final decays split primarily
between the modes $\neutralino\to\gamma+\gravitino$ and
$\neutralino\to Z + \gravitino$, and thus a preponderance of final states with
a single photon accompanied by multiple jets and \met.
To provide a signature advantageous for the photon+jets analysis,
the values of $\mu$ and $M_1$
are chosen so that, to within $\sim$ 1\%, the $\neutralino$ branching fractions are 
$B(\neutralino \rightarrow \gamma \gravitino)$ $\sim$ 50\%,
$B(\neutralino \rightarrow Z \gravitino)$ $\sim$ 49\% and
$B(\neutralino \rightarrow h \gravitino)$ $\sim$ 1\%,
irrespective of the mass of the \neutralino neutralino
($h$ represents the scalar state observed at \unit[125]{GeV},
assumed here to be the lightest CP-even state of the SUSY Higgs
spectrum). Although not explored here, the choice $\mu < 0$ would
lead to decays that prefer the production of the $h$ boson over
the $Z$ boson, producing decays rich in $b$-quark jets but otherwise similar
to the $\mu > 0$ case.

\begin{figure}[!ht]
  \begin{center}
    \includegraphics[width=0.45\textwidth]{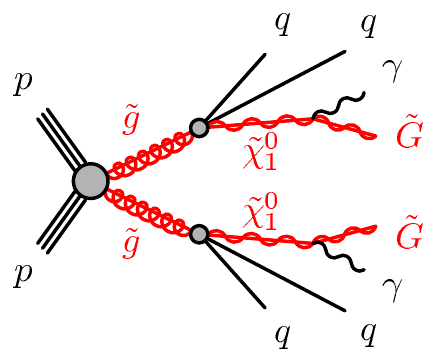} ~~
    \includegraphics[width=0.45\textwidth]{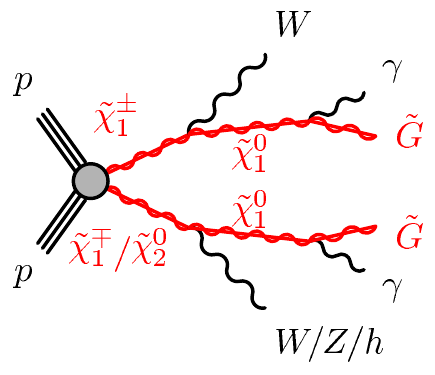}
  \end{center}
  \caption{
Typical production and decay processes for the (left) gluino-production
      and (right) electroweak-production instances of the
      GGM model for which the NLSP is a binolike neutralino. These models are
referred to in the text as the gluino--bino and wino--bino models, respectively.
    \label{fig:feynman_bino}
  }
\end{figure}

\begin{figure}[!ht]
  \begin{center}
    \includegraphics[width=0.45\textwidth]{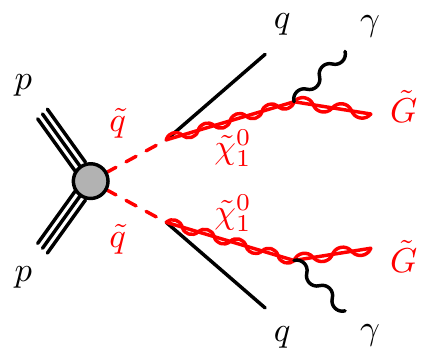}
    \includegraphics[width=0.45\textwidth]{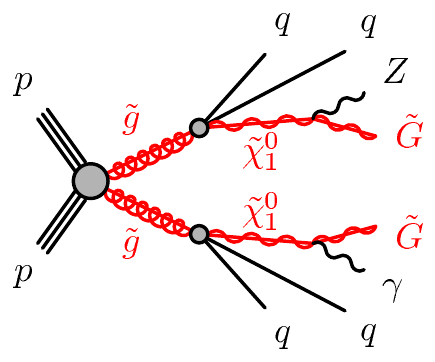}
  \end{center}
  \caption{
    Typical production and decay processes for (left) the 
      squark-production instance of the
      GGM model for which the NLSP is a binolike neutralino, and (right) the gluino-production
      instance of the
      GGM model for which the NLSP is a higgsino--bino neutralino admixture.
These models are
referred to in the text as the squark--bino and higgsino--bino models, respectively.
    \label{fig:feynman_higgsino_bino}
   }
\end{figure}

The results of the diphoton and photon+jets analyses
are interpreted in the context 
of four distinct GGM models. Three of the GGM models are associated with the
diphoton analysis, each featuring a purely binolike NLSP and distinguished by the
state directly produced by the proton--proton collision.
For the first of the three GGM models associated
with the diphoton analysis, referred to as the ``gluino--bino'' model, production proceeds through a degenerate octet of
gluinos, collectively denoted by \gluino (Figure~\ref{fig:feynman_bino} left). For the second of these models 
(the ``wino--bino'' model; Figure~\ref{fig:feynman_bino} right),
production proceeds through a degenerate triplet of the SU(2) gauge partner (wino, or \wino) states
\neutralinotwo and \chargino, and is dominated by the production
of \chiplus\chiminus and \neutralinotwo\chargino.
For the third of these models 
(the ``squark--bino'' model; Figure~\ref{fig:feynman_higgsino_bino} left), production proceeds through 
the squark states.\footnote{\StopDecFootnote} All squark states are taken to be degenerate in mass, 
with the exception of the partners of the three right-handed up-type quarks, 
whose masses are decoupled (set to inaccessibly large values) in order to satisfy GGM sum rules~\cite{Meade:2008wd}.
For a binolike NLSP, the cross section for direct \neutralino pair production
is essentially zero for any value of the \neutralino mass.
For the ``higgsino--bino'' GGM model associated with the photon+jets analysis (Figure~\ref{fig:feynman_higgsino_bino} right), 
for which the NLSP is chosen to be
a mixture of the bino and higgsino, production again proceeds through a degenerate octet of gluino states. In 
this last case, however, there is a leading-order coupling between initial-state partons and the higgsino component
of the \neutralino neutralino, leading to a SUSY production process dominated by 
\neutralino pair production for low values of the \neutralino neutralino mass. 
However, the efficiency for detecting such events in the photon+jets analysis is very small, and so 
direct \neutralino pair production is expected to play no role in the analysis. 

For all four GGM models, the masses of both the NLSP and the directly produced states are taken to be
free parameters of the model, with all other SUSY partner masses other than those of the gravitino
and $h$ state decoupled. 
The lifetime $\tau_{\neutralino}$ of the NLSP is set so that $c \tau_{\neutralino}$ is never greater than 0.1 mm. This ensures that all
particles arising from the decay of the NLSP are prompt, and in particular
that the relationship between the direction and the point of impact on
the face of the calorimeter of photons from NLSP decay is
consistent with that of a prompt photon (a separate analysis~\cite{SUSY-2013-17}
searches for GGM models with a longer-lived binolike NLSP, leading to signatures with nonprompt photons).

\section{ATLAS detector}
\label{sec:detector}

\newcommand{\AtlasCoordFootnote}{%
ATLAS uses a right-handed coordinate system with its origin at the nominal interaction point (IP)
in the center of the detector and the $z$-axis along the beam pipe.
The $x$-axis points from the IP to the center of the LHC ring,
and the $y$-axis points upwards.
Cylindrical coordinates $(r,\phi)$ are used in the transverse plane,
$\phi$ being the azimuthal angle measured relative to the $x$-axis.
The pseudorapidity is defined in terms of the polar angle $\theta$ as $\eta = -\ln [\tan(\theta/2)]$.
Angular distance is measured in units of $\Delta R \equiv \sqrt{(\Delta\eta)^2 + (\Delta\phi)^2}$.
A related quantity, $\Delta R_y$, makes use of rapidity $y$ rather than
pseudorapidity $\eta$ to define phase-space separation:
$\Delta R_y \equiv \sqrt{(\Delta y)^2 + (\Delta\phi)^2}$.}

The ATLAS detector \cite{PERF-2007-01} consists of an inner tracking
system surrounded by a superconducting solenoid, electromagnetic (EM)
and hadronic sampling calorimeters, and a muon spectrometer. The
inner detector is immersed in a 2 T axial magnetic field, and
consists of pixel and silicon microstrip detectors inside a
transition radiation tracker, providing charged-particle
tracking in the region $|\eta| < 2.5$.\footnote{\AtlasCoordFootnote}
For the $\sqrt{s}$ = 13 \tev\ run, a new innermost layer of the pixel
detector, the ``insertable B-layer''~\cite{ATLAS-TDR-19}, was added at an
average radius of $33$ mm.
The EM calorimeter uses lead as the absorber
and liquid argon (LAr) as the active material. In the central rapidity
region $|\eta| \lessapprox 1.5$, the EM calorimeter is divided into three 
layers longitudinal in shower depth,
one of them segmented into very narrow $\eta$ strips for optimal
$\gamma$/$\pi^{0}$ separation. 
The EM calorimeter is augmented by a presampler layer for
$|\eta| < 1.8$. Hadron calorimetry is based on different detector
technologies, with scintillator tiles ($|\eta| < 1.7$) or LAr ($1.5 <
|\eta| < 4.9$) as the active medium, and with steel, copper, or
tungsten as the absorber material. The muon spectrometer consists of superconducting
air-core toroids, a system of trigger chambers covering the range
$|\eta| < 2.4$, and high-precision tracking chambers allowing muon
momentum measurements for $|\eta| < 2.7$. ATLAS uses a two-level
trigger system to select events \cite{TRIG-2016-01}. A
low-level hardware trigger is implemented in custom electronics and
reduces the data rate to a design value of $\sim$ 100 kHz
using a subset of detector information. A high-level software trigger
selects events with interesting final states using software algorithms
that access the full detector information, reducing the average accepted event
rate to  $\sim$ 1 kHz.


\section{Samples of simulated processes}
\label{sec:mc}
Samples of simulated events for various $pp$ collision processes are
used to estimate the signal efficiency, develop and optimize the signal region (SR) selection,
and in some cases estimate SM background contributions to the SRs.
For the GGM model used to interpret the photon+jets results, 
the SUSY mass spectra and branching fractions are calculated using
\Suspect~2.43~\cite{Djouadi:2002ze} and
\Sdecay~1.5~\cite{Muhlleitner:2003vg}, respectively, inside the package \Susyhit~1.5a~\cite{Djouadi:2006bz},
and with Higgs boson decay provided by \Hdecay 3.4~\cite{Djouadi:1997yw}.
For the GGM models used to interpret the diphoton results,
the SUSY mass spectra and branching fractions are calculated using
\Suspect~2.41~\cite{Djouadi:2002ze} and
\Sdecay~1.3b~\cite{Muhlleitner:2003vg}, respectively.
For all models, the Monte Carlo~(MC) SUSY signal samples were generated to leading-order accuracy using
\texttt{MG5$_{-}$aMC@NLO} v2.3.3~\cite{Alwall:2014hca}, with up to two extra 
partons included beyond the underlying $2 \rightarrow 2$ SUSY production process.
The simulation used
the \texttt{NNPDF2.3LO} parton distribution functions (PDF) 
set~\cite{Ball:2012cx}, and was interfaced to \texttt{PYTHIA} 8.212~\cite{Sjostrand:2014zea}
with the ATLAS A14 set of tuned parameters~\cite{ATL-PHYS-PUB-2014-021} for the modeling
of the parton showering, hadronization and underlying event.
Strong and electroweak SUSY production cross sections are calculated to next-to-leading order (NLO) 
in the strong coupling constant, adding the 
resummation of soft gluon emission at next-to-leading-logarithm accuracy 
(NLO+NLL)~\cite{Beenakker:1996ch,Kulesza:2008jb,Kulesza:2009kq,Beenakker:2009ha,Beenakker:2011fu,Fuks:2012qx,Fuks:2013vua}. The nominal 
cross section and its uncertainty are taken from an envelope of cross-section predictions using different PDF 
sets and factorization and renormalization scales, as described in Ref.~\cite{Borschensky:2014cia}.

While most of the backgrounds to the GGM models under examination
are estimated through the use of control samples selected from data, as described below,
the extrapolation from control regions (CRs) to signal regions depends on samples of simulated events,
as do the optimization studies.
Simulated SM processes include single-photon and diphoton production both with and without an associated vector boson,
\ttbar production both with and without an accompanying photon, and multijet production.
With the exception of the $\ttbar\gamma$ process,
Standard Model processes were generated using the \texttt{SHERPA v2.1.1} simulation package~\cite{Gleisberg:2008ta},
making use of the \texttt{CT10}~\cite{Lai:2010vv} PDF set. Matrix elements were calculated for up to three-parton 
emission at leading order (LO) using the \texttt{COMIX}~\cite{Gleisberg:2008fv} generator and then combined with the \texttt{SHERPA} 
parton shower~\cite{Schumann:2007mg} according to an improved CKKW procedure~\cite{Hoeche:2009rj}.
The $\ttbar\gamma$ process was generated to next-to-leading-order accuracy using 
\texttt{MG5$_{-}$aMC@NLO} v2.3.3~\cite{Alwall:2014hca} in conjunction 
with \texttt{PYTHIA} 8.186~\cite{Sjostrand:2007gs}
with the \texttt{NNPDF2.3LO} PDF set and the A14 set of 
tuned parameters.

All MC samples were processed with the \texttt{GEANT4}-based
simulation~\cite{SOFT-2010-01,Agostinelli:2002hh} of the ATLAS
detector, or, where appropriate, a simulation of the
ATLAS detector based on parametrized shower shapes in the
calorimeter, and \texttt{GEANT4} elsewhere. Corrections are applied to
the samples of simulated events to account for differences between data and
simulation in the photon-based trigger, identification, and
reconstruction efficiencies, as well as for the efficiency and
misidentification rate of the algorithm used to identify jets
containing $b$-hadrons ($b$-tagging).  
The effect of additional $pp$ interactions per bunch crossing (``pileup'') is taken into
account by overlaying
simulated
minimum-bias events according to the observed distribution of the
number of pileup interactions in data.

\section{Reconstruction of candidates and observables}
\label{sec:objects}

Primary vertices are formed from sets of two or more tracks,
each with transverse momentum $p_{\mathrm T} >$ \unit[400]{MeV}, that are
consistent with having originated at the same three-dimensional
space point within the luminous region of the colliding proton beams.
When more than one such
primary vertex is found, the vertex with the largest scalar sum
of the squared transverse momenta of the associated tracks is chosen.

Electron
candidates are reconstructed from EM calorimeter energy clusters
consistent with having arisen from the impact of an electromagnetic
particle (electron or photon) upon the face of the calorimeter.
For the object to be considered an electron, it is required to
match a track reconstructed by an algorithm
optimized for recognizing charged particles with a high probability of bremsstrahlung.
Electrons are required to pass a ``tight'' set
of identification requirements as defined in Refs.~\cite{PERF-2016-01,PERF-2013-05,ATLAS-CONF-2016-024},
based on the characteristics of the EM shower development, the quality of the 
associated reconstructed track, and the quality of the association of the track with the 
calorimeter deposition. 
Electron
candidates used by these searches are further required to have
$\pt > \unit[25]{GeV}$ and $|\eta| < 2.47$, but excluding the
transition region $1.37 < \left|\eta\right| < 1.52$ between the barrel
and end cap calorimeters. A track-based isolation requirement is
imposed, with the scalar sum of the transverse momenta of tracks within a cone
of size $\Delta R = 0.2$ (excluding that of the electron candidate's track) 
required to be less than a value that leads to
a loss of efficiency of 5\% for electrons with $\pt = 25$~GeV, and
of less than 1\% for electrons with $\pt > 60$~GeV.
Finally, the electron track is required to be consistent with
having originated from the primary vertex in the $r$--$z$ plane.

Electromagnetic clusters in the range $\left|\eta\right| < 2.37$ (excluding the transition region
$1.37 < \left|\eta\right| < 1.52$)
are classified as photon candidates provided
that they either have no matched track (``unconverted'' photons)
or have
one or more matched tracks consistent with having originated from a photon conversion vertex (``converted'' photons).
Photon candidates are required to have $\Etg > 25$~\GeV, where \Etg is the
energy of the photon candidate, measured in the EM calorimeter, multiplied
by the cosine of the angle of its trajectory relative to the plane perpendicular to the $z$ axis.
The photon direction is estimated either
using EM calorimeter shower-depth segmentation (if unconverted) or the position of the conversion
vertex (if converted), together with constraints from the $pp$ collision point.
Photon candidates are also required to fulfill
``loose'' or ``tight'' 
identification criteria~\cite{ATL-PHYS-PUB-2011-007,PERF-2013-04} based on observables that reflect the shape of
the electromagnetic showers in the calorimeter, in particular in the finely segmented
first layer. While tight photons are required for all SRs, loose photons are used to
construct control samples that aid in the estimation of backgrounds arising from
misreconstructed jets. 
If an EM calorimeter
deposition is identified as both a photon and an electron, the photon
candidate is discarded and the electron candidate retained.
Additionally, a calorimeter-based isolation requirement is imposed:  after correcting for
contributions from pileup and the deposition ascribed to the photon
itself, the transverse energy \EtOFour deposited in a cone of size $\Delta R = 0.4$
surrounding the photon candidate's energy deposition must satisfy the relation
$\EtOFour <$ \unit[2.75]{GeV} + $0.22 \times \Etg$, with \Etg in GeV.

Muon candidates are reconstructed via a combination of track information from the
muon spectrometer and the inner tracking systems. 
Muons must pass the ``medium'' identification
requirements defined in Ref.~\cite{PERF-2015-10}, based on requirements on the number of hits in the different inner detector and
muon spectrometer subsystems, and on the significance of the charge-to-momentum ratio measurement.
Muon candidates are required to have 
$\pt >$ \unit[25]{GeV} and $\left|\eta\right| < 2.7$. Muon candidates are also required
to pass an isolation requirement identical to that for electron candidates.
Finally, the muon track
is required to be consistent with having originated from the primary vertex in both the
$r$--$z$ and $r$--$\phi$ planes.

Making use of utilities within the \texttt{FastJet} package~\cite{ref:FastJet}, jets are reconstructed 
from three-dimensional energy clusters in the calorimeter~\cite{PERF-2014-07-pub} 
with the anti-$k_t$ jet
clustering algorithm~\cite{Cacciari:2008gp} with a radius parameter $R$ = 0.4. 
In the diphoton analysis, only jet candidates with $\pt >$ \unit[30]{GeV} and $|\eta| < 2.8$
are considered. For jets used in the photon+jets analysis, the acceptance
is further reduced to $|\eta| < 2.5$.
Jets are calibrated as described in Refs.~\cite{PERF-2015-05,PERF-2016-04},
with the expected average energy
contribution from pileup clusters subtracted in accordance with the angular area of the jet.
Jets resulting from the hadronization of $b$-quarks are identified using
the multivariate \texttt{MV2c10} $b$-tagging algorithm, which is based on quantities such as impact parameters of
associated tracks, and reconstructed secondary vertices~\cite{ATL-PHYS-PUB-2016-012,PERF-2012-04}. 
This algorithm is used at a working point
that provides 77\% $b$-tagging efficiency in simulated \ttbar events, and a rejection factor of 134 for light-quark
and gluon jets and 6 for charm jets.

To avoid ambiguity that arises when an electron or photon is also reconstructed as a jet,
the following procedure is used:
if a jet and an electron or photon are reconstructed with a
separation of
$\Delta R_y < 0.2$, the electron or photon is retained and
the jet is discarded; if $0.2 < \Delta R_y < 0.4$
then the jet is retained and the electron or photon is discarded.
Finally, in order to suppress the reconstruction of muons arising from showers induced
by jets, if a jet and a muon are found with $\Delta R_y < 0.4$ the jet is retained
and the muon is discarded.

The vector momentum imbalance \ptm in the transverse plane is obtained from the negative vector
sum of the reconstructed and calibrated physics objects, and an additional soft term. The soft term 
is constructed from all tracks that are not associated with
any reconstructed electron, muon or jet, but which are associated with the 
primary vertex.

Several additional observables are defined to help in the discrimination of SM backgrounds
from potential GGM signals. 
The ``effective mass'' \MEFF
is defined as the scalar sum of the transverse energy of identified photons,
any additional leptons and jets in the event, plus the value of \met.
The ``photon-enhanced'' total visible transverse energy observable \HT is defined
as the transverse energy of the selected photons and any additional leptons and
jets in the event, without the addition of \met. In this case the contribution from photonic signatures is
emphasized by discarding the
photon-jet ambiguity resolution procedure when identifying photons and jets.
Requiring a minimum
value for either of these observables exploits the high energy scale
associated with the production of massive SUSY partners.
The photon--\met separation
\dphi is defined as the azimuthal
angle between the \ptm vector and the 
selected photon. In the diphoton analysis, \dphim is
defined to be the minimum value of \dphi of the two
selected photons. The minimum jet--\met separation \dphijm
is defined as the minimum azimuthal angle between
the \ptm vector
and the two leading (highest-\pt)
jets in the event.
For the diphoton analysis, leading jets are required
to have $\pt >$ \unit[75]{GeV} for the purpose of constructing this
observable, and if no such jet is found no requirement is placed on
the observable. 
Small values of these angular-separation observables are often
associated with SM backgrounds arising from poorly reconstructed photons or jets.
Finally, the quantity \RT is defined as
the scalar sum of the transverse momenta of the four highest-\pt
jets in the event divided by the scalar sum of the
transverse momenta of all jets in the event; smaller values of \RT
are typical for the jet-rich events of the higgsino--bino GGM model
that is the focus of the photon+jets analysis.

\section{Event selection}
\label{sec:selection}
The data sample
is selected by a trigger
requiring the presence of one loose photon
with $\ET > 140$~GeV for the
photon+jets analysis, or two loose photons with
$\ET > 35$~GeV and $\ET > 25$~GeV, respectively, for the diphoton analysis.
After applying data-quality requirements related to the beam and detector conditions, the total
available integrated luminosity is \integLumi.

For the diphoton analysis, targeting the exploration of
the gluino--bino, squark--bino and wino--bino GGM models incorporating a purely binolike \neutralino,
two separate SR selection strategies
are used: a ``\BS'' selection targeting the production
of higher-mass strongly coupled SUSY states (gluinos and squarks) and a
``\BW'' selection targeting the production of lower-mass
weakly coupled SUSY states (winos). For each of these approaches,
two SRs are defined: the first (\BSL, \BWL) optimized for the
case of a lower-mass \neutralino and the second (\BSH, \BWH) for
a higher-mass \neutralino. For fixed production-scale (gluino, squark, wino) mass, increasing the
mass of the bino NLSP increases the energy carried off by the unobserved gravitinos,
at the expense
of the overall visible energy deposition.

For the photon+jets analysis, targeting the higgsino--bino
GGM model,
a further two SRs are defined.
The first of these (\HBL) is optimized for a high-mass gluino and a low-to-intermediate
mass neutralino, for which there is a large mass difference between the gluino and the neutralino.
Such events are characterized by
large jet multiplicity and exceptional hadronic activity, but moderate missing
transverse momentum.
The second of these SRs (\HBH)
targets the compressed scenario for which the difference between the
gluino and neutralino masses is small,
resulting in lower jet multiplicity and suppressed hadronic activity while producing harder photons and
greater missing transverse momentum.

All four diphoton SRs require two tight, isolated photons with $\ET > \unit[75]{GeV}$,
while \HBL and \HBH require a single tight, isolated photon with $\ET > \unit[145]{GeV}$
and $\ET > \unit[400]{GeV}$, respectively.
To exploit the transverse momentum imbalance created by the unobservable gravitinos,
an event must exhibit significant \met to be included in any of the SRs.
To ensure that the \met observable is accurately measured, minimum requirements
on \dphim and \dphijm are considered for each SR.

\begin{figure}[!htb]
  \begin{center}
    \includegraphics[width=0.48\textwidth]{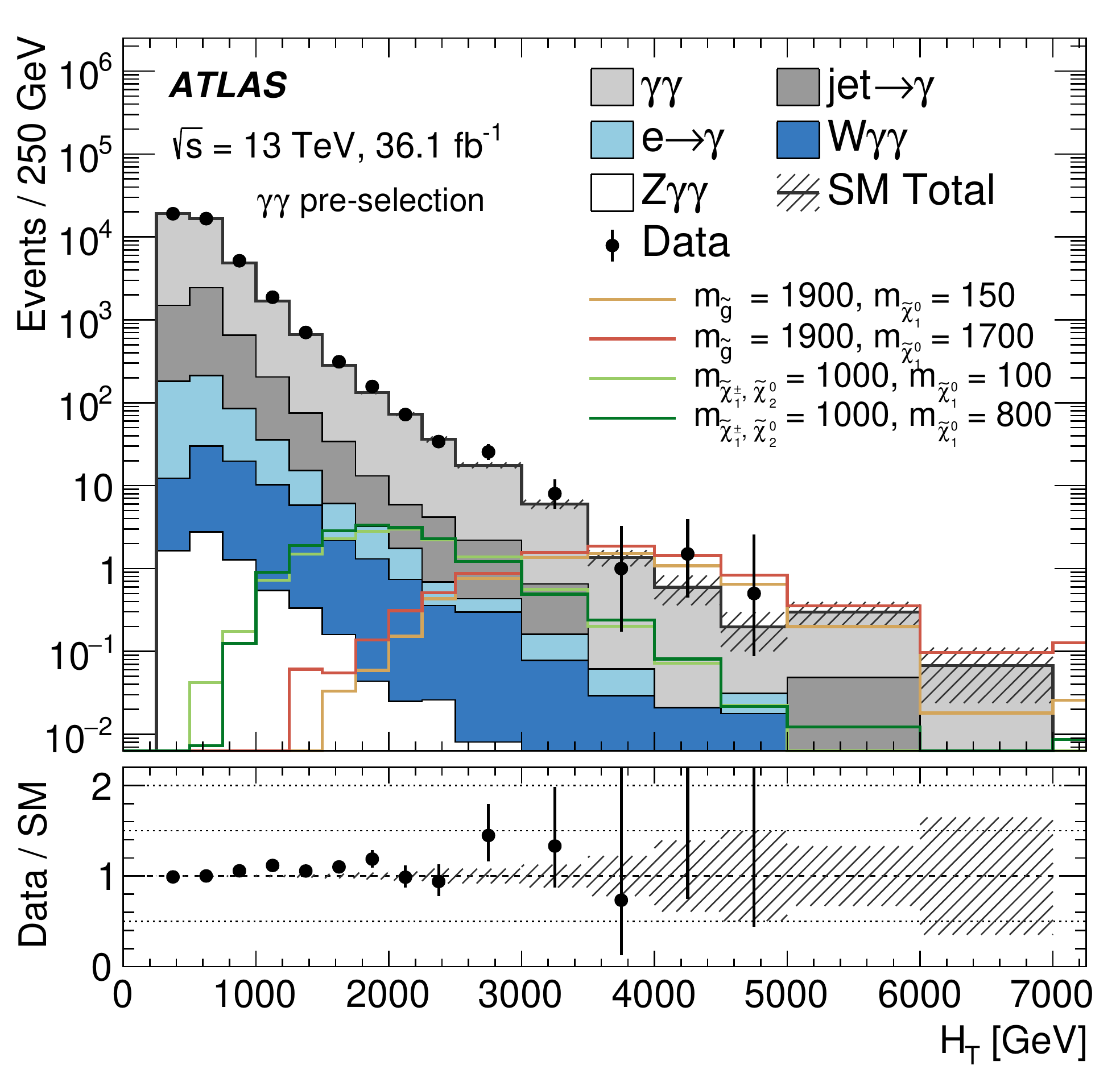}
    \includegraphics[width=0.48\textwidth]{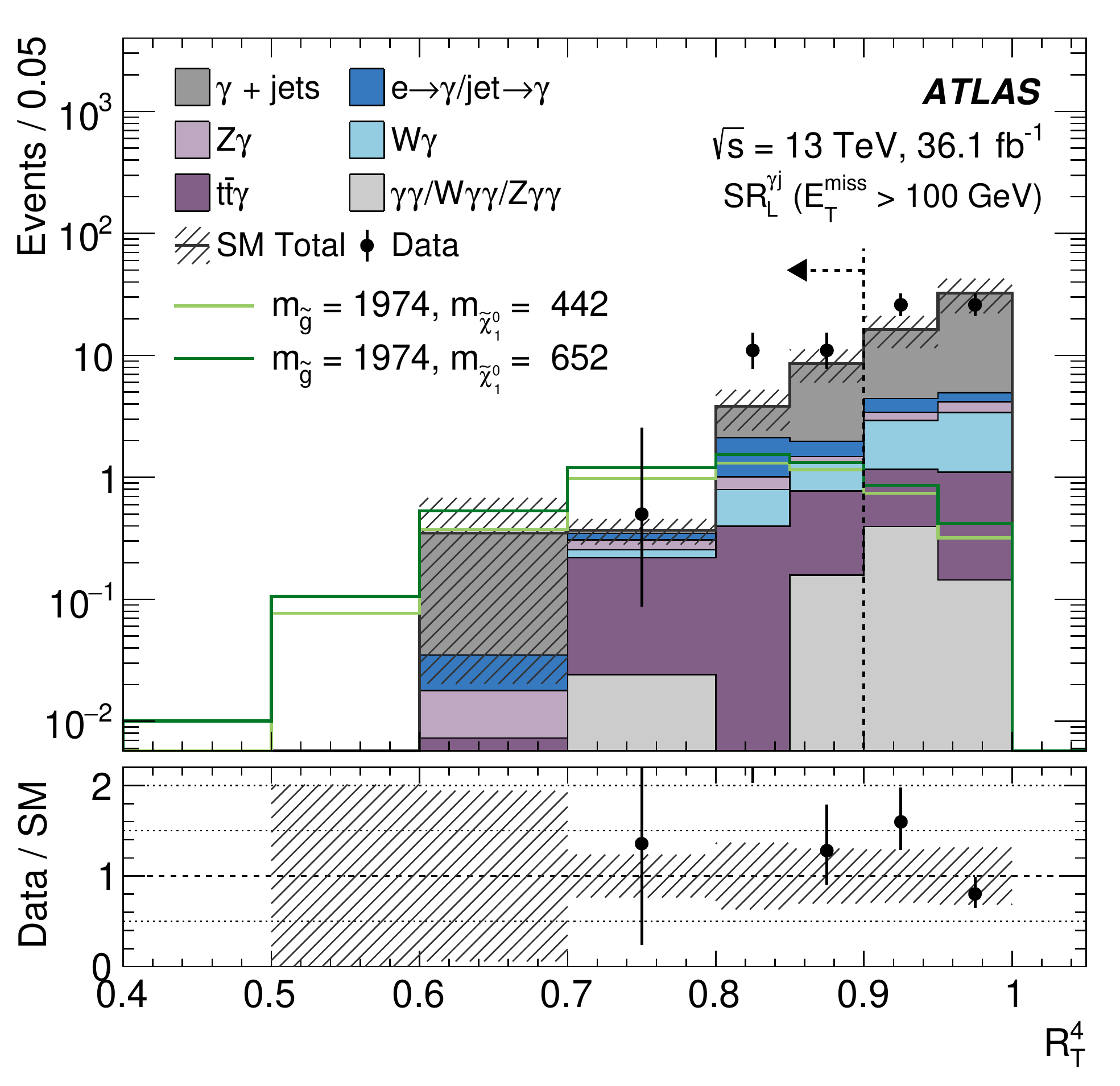}
  \end{center}
  \caption{
    Left: distribution of the total visible transverse energy \HT
    for selected diphoton events, after requiring $\dphijm > 0.5$ but before
    application of a requirement on \MET and \dphim (``$\gamma \gamma$ preselection''). 
    Also shown are the expected \HT distributions of contributing SM processes as 
    well as those for two points each in the 
    parameter spaces of the gluino--bino and wino--bino GGM models (mass values in GeV).
    Events outside the range of the displayed region are included in the highest-value bin.
    Right: distribution of \RT
    for the sample satisfying all \HBL selection
    criteria except the \RT requirement itself, but with a relaxed requirement
    of $\MET > 100$~GeV. 
    Also shown are the expected \RT distributions of contributing SM processes as well as 
    those for two points in the $m_{\gluino}$--$m_{\neutralino}$ parameter space of the
    GGM model relevant to the photon+jets analysis (mass values in GeV). The value of the gluino mass arises from the
    choice $M_3 = 1900$~GeV, while the values of the \neutralino mass arise from the 
    choices $\mu = 400$ and $\mu = 600$~GeV, combined with the constraint
    that the branching fraction of $\neutralino \rightarrow \gamma \gravitino$ be 50\%.
The vertical dashed line and left-pointing arrow show the region of the
\RT observable selected for inclusion in \HBL.
Uncertainties are shown as hatched bands for the various expected
sources of SM background (statistical only) and as error bars for
data. The lower panels show the ratio of the data to the SM prediction.
    }
    \label{fig:diphoton_meff}
\end{figure}

Requirements are
made on a number of additional observables, defined in Section~\ref{sec:objects}, with values chosen to optimize
the sensitivity to the GGM signal of interest in each SR.
To exploit the
high energy scale associated with SUSY production at masses close to
the expected limit of sensitivity of the various SRs, all SRs include minimum requirements
on one of the two total-transverse-energy observables \HT or \MEFF.
As an illustration,
Figure~\ref{fig:diphoton_meff}~(left) shows the \HT distribution of diphoton
events as well as that expected from SM sources (estimated as described in Section~\ref{sec:background})
and from four
characteristic scenarios of the binolike NLSP GGM gluino-production model.
Due to the large backgrounds arising from SM single-photon production, requirements must be placed on additional
observables in order to optimize the signal sensitivity in the photon+jets analysis.
A minimum of five (three) jets is required for events in \HBL (\HBH).
For \HBL of the photon+jets analysis, an additional
requirement that
events have $\RT < 0.90$ helps reduce the background from SM events,
which tend to have fewer and softer jets 
than do signal events. Examples of the discriminating power of the \RT 
observable are shown in Figure 3 (right).
Finally, for both \HBL and \HBH,
events with one or more leptons (electron or muon) are rejected in order to suppress
the contribution from SM events containing leptonically decaying $W$ or $Z$ bosons
produced in association with a hard radiated photon (``$V\gamma$'' production).
In addition, a predecessor to \HBL, originally designed for a search using a
smaller data set (13.2 fb$^{-1}$), has been retained, as the number of events
observed in that search exceeded the
background prediction. This third photon+jets SR is referred to as \HBLP,
and differs from \HBL only by the relaxed requirement $\MET > 200$~GeV
relative to the $\MET > 300$~GeV requirement of \HBL.
A summary of the selection requirements for the various SRs is presented in Table~\ref{tab:selection}.

\begin{table}[t]
  \caption{The requirements
defining the seven SRs for the diphoton and photon+jets
searches. All symbols are defined in the text.
An ellipsis is entered when no such requirement is
made in the given signal region.
}
  \begin{center}
    \sisetup{retain-explicit-plus}
  \begin{tabular*}{\textwidth}{@{\extracolsep{\fill}}lrrrrrrr} \hline\hline\noalign{\smallskip}
    {\textbf Signal region}   &    \BSL  &    \BSH  & \BWL     & \BWH     &  \HBL    &  \HBLP    & \HBH      \\
      \noalign{\smallskip}\hline\noalign{\smallskip}
    Number of photons         &$\ge 2$   &$\ge 2$   &$\ge 2$   &$\ge 2$   &$\ge 1$   &$\ge 1$    &$\ge 1$    \\
    \Etg [GeV]                & $>75$    & $>75$    & $>75$    & $>75$    & $>145$   & $>145$    &  $>400$   \\
    Number of jets            & ...      & ...      & ...      & ...      & $\ge 5$  & $\ge 5$   & $\ge 3$   \\
    Number of leptons         & ...      & ...      & ...      & ...      & 0        & 0         &  0        \\
    $\met $ [GeV]             & $> 150$  & $> 250$  & $> 150$  & $> 250$  & $> 300$  & $> 200$   & $> 400$   \\
    $\HT$  [GeV]              & $>2750$  & $>2000$  & $>1500$  & $>1000$  & ...      & ...       & ...       \\
    $\MEFF$ [GeV]             & ...      & ...      & ...      & ...      & $>2000$  & $>2000$   & $>2400$   \\
    $\RT$                     & ...      & ...      & ...      & ...      & $< 0.90$ & $< 0.90$  & ...       \\
    $\dphijm$                 & $> 0.5$  & $> 0.5$  & $> 0.5$  & $> 0.5$  & $> 0.4$  & $> 0.4$   & $> 0.4$   \\
    $\dphim$ ($\dphi$)        &    ...   & $> 0.5$  & ...      & $> 0.5$  & ($> 0.4$)& ($> 0.4$) & ($> 0.4$) \\
      \noalign{\smallskip}\hline\hline\noalign{\smallskip}
  \end{tabular*}
  \label{tab:selection}
  \end{center}
\end{table}

\section{Background estimation}
\label{sec:background}
Backgrounds to the various SRs arise from a number of sources
that generate real photons in combination with energetic neutrinos,
as well as events in which one or more energetic jets or electrons are misidentified as photons.
In the following, the methodology of the background estimation for the two experimental
signatures is discussed, and the resulting background estimates, broken down by source, are tabulated.
Backgrounds arising from misidentified jets and electrons are estimated through the use of control
samples including jets or electrons, scaled by misidentification rates determined from data. Other backgrounds are
estimated via MC simulation, often constrained by observed event counts in dedicated CRs.
For the estimation of background contributions that rely upon MC simulation, either directly
or through the estimation of ``transfer factors'' relating the background content of CRs to
that of corresponding SRs, the effect of MC modeling uncertainties
is considered.

In the photon+jets analysis, expected SM backgrounds constrained by CRs
are determined separately for each SR with a maximum-likelihood fit,
referred to as the ``background-only fit.'' The background-only fit constrains the normalization of the
dominant backgrounds to the observed event yields in the
associated CRs, assuming that no signal is present in the CRs.
The inputs to the fit for each SR include the numbers of events observed in its associated CRs and the
number of events predicted by simulation in each region for all background processes. The latter are
described by Poisson statistics. The systematic uncertainties in the expected values are included in the
fit as nuisance parameters, modeled by Gaussian distributions with widths corresponding to
the sizes of the associated uncertainties. Correlations between the various CRs are taken into account.
The product of the various probability density functions forms the likelihood, which the fit maximizes
by adjusting the background normalization and the nuisance parameters.
Background models are confirmed in validation regions (VRs) with
selection criteria closely related to those of the corresponding SR, but with one or more selection
criteria modified to suppress the potential contribution of a GGM signal to the VR.

\subsection{Backgrounds to the diphoton analysis}
\label{sec:sub_binoback}

Backgrounds from SM contributions to the four diphoton SRs
are grouped
into three primary components.  The first of these, referred
to as ``QCD background,'' arises from a mixture of
processes that include $\gamma \gamma$ production as well as
$\gamma$ + jet and multijet events with at least one jet misreconstructed as a
photon.  The second background component, referred to as ``EW background,'' is due
primarily to $W+X$
(here ``$X$'' can be any number of jets, accompanied
by no more than one photon; the two-photon case is treated separately)
and \ttbar events.
These events tend to include final-state neutrinos that produce
significant \MET. In both cases, EW background events entering the signal regions
generally have at least one electron misreconstructed as a photon.
The QCD and EW backgrounds
are estimated through the use of dedicated control samples of data events.

The third background component,
referred to as ``irreducible,''
consists of $W$ and $Z$ bosons produced in association with two
real photons, with a subsequent decay into one or more neutrinos.
For this background, the $\Wgg$ component dominates, and
requires corrections to its LO contribution that
are both large and rapidly varying across the phase space of
the $\Wgg$ (plus possible additional jets) process~\cite{Bozzi:2011wwa}. Thus
a data-driven approach is developed to constrain the
$\Wgg$ contribution to the four SRs. The \Zgg contribution is estimated directly from the MC simulation.

The QCD background to \BSL, \BSH, \BWL and \BWH
is expected to arise from events with a single real, isolated photon and a jet whose
fragmentation fluctuates in such a manner as to cause it to be misidentified
as a second isolated photon (``jet $\rightarrow \gamma$'' events), and, to a lesser extent,
from events with two real, isolated photons
unaccompanied by any additional electroweak bosons (``QCD diphoton'' events).
The contribution from
dijet events is found to be small and largely incorporated into the
jet $\rightarrow \gamma$ background estimate.

To estimate the jet $\rightarrow \gamma$ contribution,
a ``\QCDg control sample'' is identified
within the diphoton-trigger data sample by selecting events for which one photon candidate satisfies
the tight selection criterion, while the other
satisfies the loose but not the tight photon
criterion. Both photons are required to have $\Etg > 75$~GeV, and
events containing electrons are vetoed to reduce contamination from
$W\to e\nu$ decays.
A model of the jet $\rightarrow \gamma$ background is then obtained by
multiplying the number of control-sample events by a loose-to-tight scale
factor in the range 0.1--0.5, depending upon the values of \pt and $\eta$ of the loose photon,
determined from events with poorly isolated photons ($10 < \EtOFour - 0.22 \times \Etg < 30$~GeV). Studies
with MC simulated samples as well as \MET and \HT sideband data
show this sample to be dominated by misreconstructed particles in hadronic jets, and also
suggest that the \MET distribution of this control sample adequately
reproduces the \MET distribution of the QCD background in the high-\MET
region used for the signal selection.

A diphoton MC sample, scaled as a function of \MET and the number of jets to reproduce the
observed numbers of data events in the region  $0 < \MET < 150$~GeV, is used for the estimation of the small
diphoton contribution to the QCD background. Before the application of a requirement on \HT, and for each bin in
the number of observed jets, an \MET-dependent
scale factor of between 0.7 and 1.3 is applied to the MC simulation to establish agreement between data and simulation. The scaling behavior for
values of \MET in the diphoton SRs is estimated by extrapolating the \MET dependences of the scale factors
observed for $\MET < 150$~GeV into the region $\MET > 150$~GeV. This procedure yields the level of agreement
between the data and MC distributions of \HT illustrated in Figure~\ref{fig:diphoton_meff}.

For each SR, the jet $\rightarrow \gamma$ (QCD diphoton) background estimate is obtained by counting the number of scaled
\QCDg control (diphoton MC) events satisfying the combined \MET, \HT and $\Delta \phi$ requirements for the given SR.
The statistical uncertainty in each estimate is determined according
to the unscaled number of events in the \QCDg control and diphoton MC samples that satisfy these requirements.
If no events remain in the given sample, a one-sided statistical uncertainty is adopted,
corresponding to the $68\%$ confidence level (C.L.) Poisson upper limit on the possible background contribution.
An additional uncertainty of $\pm 50\%$ is included to account for possible
modeling uncertainties.
The resulting QCD background estimates and their overall uncertainties
are shown in Table~\ref{tab:diphoton_back}, separately for the jet $\rightarrow \gamma$ and QCD diphoton
contributions.

\begingroup
\renewcommand*{\arraystretch}{1.15}
\begin{table}[t]
    \caption{The expected and observed numbers of events for
      the four diphoton signal regions. The quoted errors are the combined
      statistical and systematic uncertainties.}
  \begin{center}
  \begin{tabular*}{\textwidth}{@{\extracolsep{\fill}}lcccc} \hline\hline\noalign{\smallskip}
      {Signal region} & \BSL & \BSH & \BWL & \BWH  \\
      \noalign{\smallskip}\hline\noalign{\smallskip}
  Jet $\rightarrow \gamma$ & $0.19^{+0.21\phantom{1}}_{-0.19\phantom{1}}$ & $0.19^{+0.21\phantom{1}}_{-0.19\phantom{1}}$ & $0.93 \pm 0.67$        & $0.19^{+0.21\phantom{1}}_{-0.19\phantom{1}}$ \\
  QCD diphoton             & $0.00^{+0.17\phantom{1}}_{-0.00\phantom{1}}$ & $0.00^{+0.17\phantom{1}}_{-0.00\phantom{1}}$ & $0.15^{+0.17\phantom{1}}_{-0.15\phantom{1}}$ & $0.00^{+0.17\phantom{1}}_{-0.00\phantom{1}}$ \\
  EW background            & $0.08 \pm 0.04$        & $0.06 \pm 0.04$        & $0.88 \pm 0.23$        & $0.51 \pm 0.15$        \\
  $(W\to\lnu)\gamma\gamma$ & $0.22 \pm 0.14$        & $0.21 \pm 0.13$        & $1.55 \pm 0.78$        & $1.08 \pm 0.56$        \\
$(Z\to\nu\nu)\gamma\gamma$ & $0.01 \pm 0.01$        & $0.03 \pm 0.02$        & $0.15 \pm 0.08$        & $0.27 \pm 0.13$        \\
\hline
Expected background events & $0.50^{+0.30\phantom{1}}_{-0.26\phantom{1}}$ & $0.48^{+0.30\phantom{1}}_{-0.25\phantom{1}}$ & $3.7 \pm 1.1$        & $2.05^{+0.65\phantom{1}}_{-0.63\phantom{1}}$ \\
\hline
     Observed events & 0 & 0 & 6 & 1 \\
 \hline\hline
   \end{tabular*}
   \label{tab:diphoton_back}
  \end{center}
\end{table}
\endgroup

The EW background is estimated via an ``electron--photon control sample''
composed of events with at least one isolated tight photon and one isolated electron, each with
$\ET > 75$~GeV; when there is more than one identified electron,
the one with the highest \pt is used.
The electron--photon control sample is scaled by the probability for such an electron to
be misreconstructed as a tight photon, as estimated from a comparison of the
rate of $Z$ boson reconstruction in the $e\gamma$ and $ee$ final states. The electron-to-photon scale factor varies
between 1\% and 5\%, with larger factors associated with larger values of $|\eta|$,
since the misidentification rate depends on the amount of material in front of the calorimeter.
Events with additional photons or leptons are vetoed from the control sample to preserve
its orthogonality to the various diphoton and photon+jets SRs.
After applying all additional selection requirements to the scaled electron--photon control sample, and
including a systematic uncertainty of $\pm 20$\% associated with the determination of the scale factor,
the resulting estimates of the EW background
to the four diphoton SRs are shown in Table~\ref{tab:diphoton_back}.

The \Wgg background to the four diphoton SRs is estimated using
a lepton--diphoton (\lgg) CR. To enhance
the contribution of \Wgg and to ensure that the \lgg CR
is exclusive of the four SRs, the photon \et requirement is lowered to
50~GeV and a requirement of $50 < \MET < 150$~GeV is imposed. To ensure that
the CR sample arises from the same region of the \Wgg process phase space
as the expected background, a further requirement that the transverse momentum
of the \lgg system be greater than 100~GeV is imposed.
A total of 13 events is observed in the CR, for which MC simulation suggests that 3.9 events
are expected to arise from SM sources other than \Wgg.
In the limit that no GGM signal contributes to the \lgg control region,
an enhancement factor of $1.6 \pm 0.6 \pm 0.4$ must be applied to the \Wgg MC sample to achieve
agreement between the MC simulation and data in the \lgg control region.
The statistical uncertainty of $\pm 0.6$ arises from the Poisson error in the difference
between the observed number of events in the \lgg control region and the number of events expected from SM processes
other than \Wgg production.
The systematic uncertainty of $\pm 0.4$ arises from assuming that the non-\Wgg
contributions to the \lgg CR have an uncertainty of 100\%; this uncertainty dominates smaller
contributions arising from potential mismodeling of the detector response.
For each diphoton SR, the \Wgg background estimate is then provided by applying
all associated SR requirements to the scaled \Wgg MC sample. 
The resulting \Wgg background estimate in each of the four SRs,
assuming that there is no signal contribution to the \lgg CR,
is shown in Table~\ref{tab:diphoton_back}.
Also shown is the combined background estimate, including uncertainty,
from all SM sources; for the \Zgg background, an uncertainty
of $\pm 45\%$ is assigned to account for the effect of QCD scale dependence
associated with the limited-order simulation
of the \Zgg process discussed in Section~\ref{sec:mc}.

\begin{figure}[!htb]
  \begin{center}
    \includegraphics[width=0.96\textwidth]{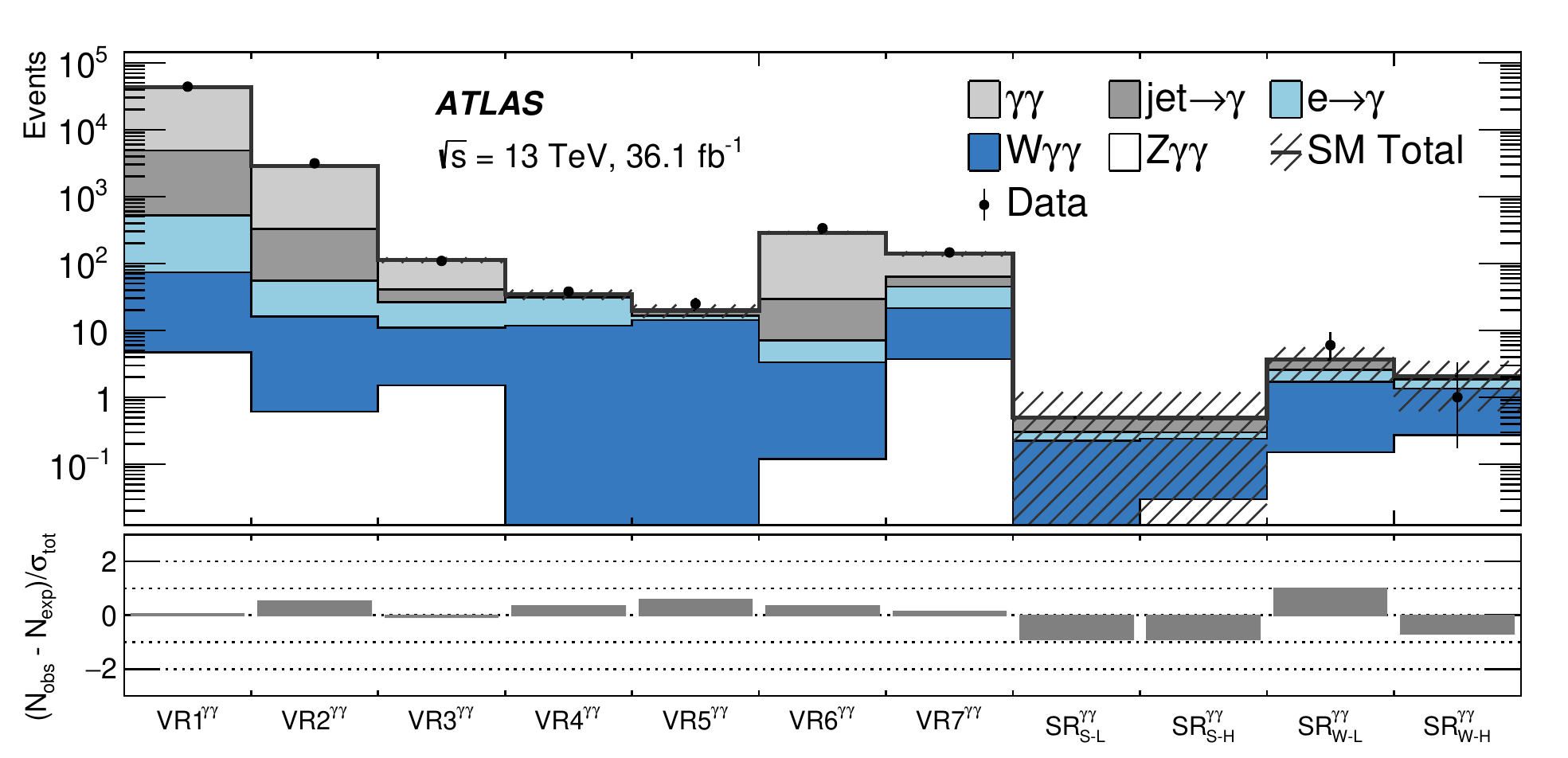}
  \end{center}
  \caption{Comparisons between expected and observed content of the
validation and signal regions for the diphoton analysis. 
The uncertainties in the numbers of expected events are
the combined statistical and systematic uncertainties.
The lower panel shows the pull (difference between observed and expected
event counts normalized by the uncertainty) for each region.
    \label{fig:diphoton_reg}
  }
\end{figure}

The accuracy of the resulting overall background model
is confirmed by the use of seven VRs that, while excluding events in
the four diphoton SRs, have kinematic properties similar to those of the signal region.
The definitions of these VRs are shown in Table~\ref{tab:diphoton_vr}, together with the expected and observed numbers of events in each region.  
Figure~\ref{fig:diphoton_reg} also shows this comparison,
with the expected number of events broken down into its contributing SM sources.

\begin{figure}[!htb]
  \begin{center}
    \includegraphics[width=0.48\textwidth]{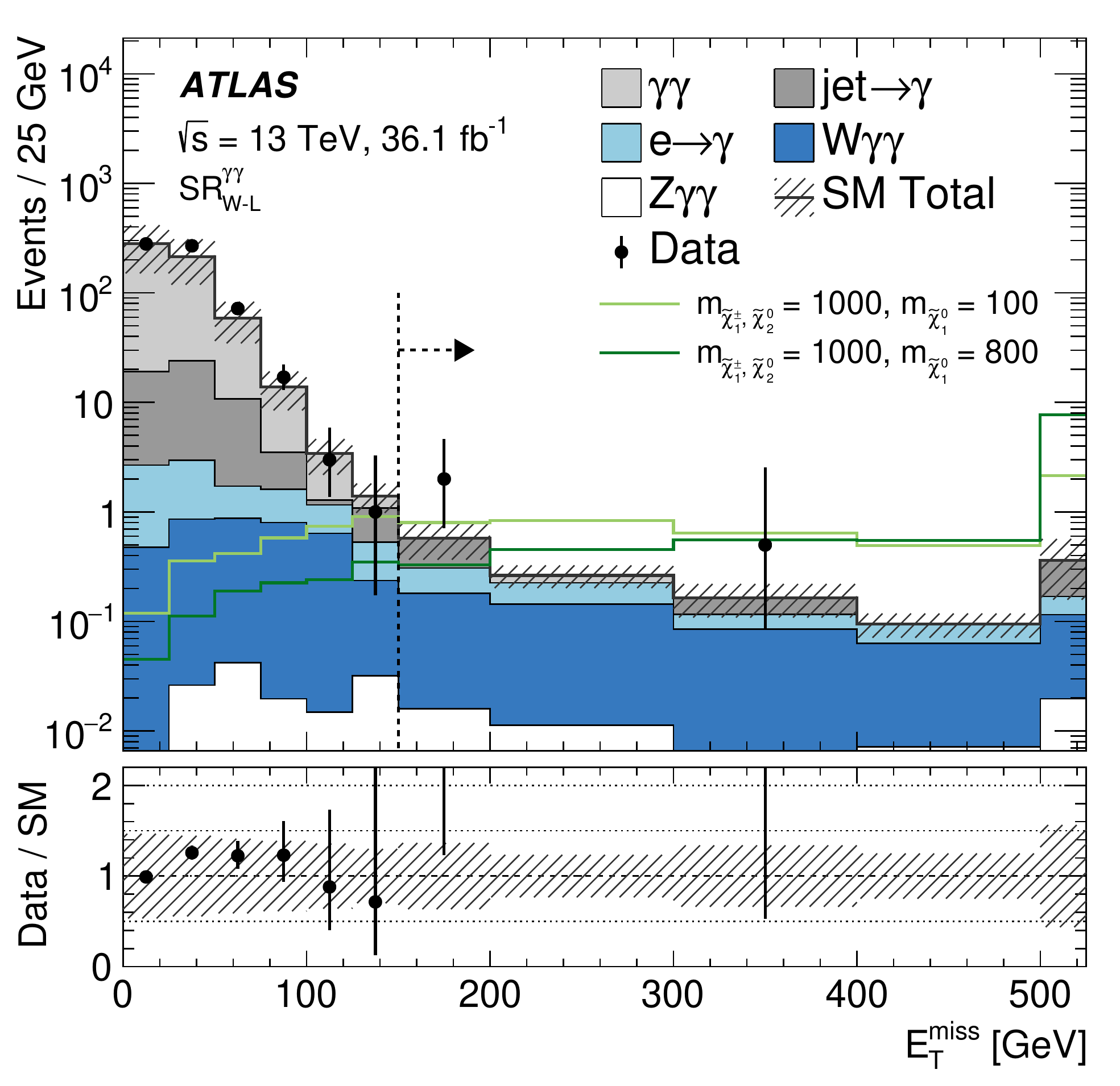} ~~
    \includegraphics[width=0.48\textwidth]{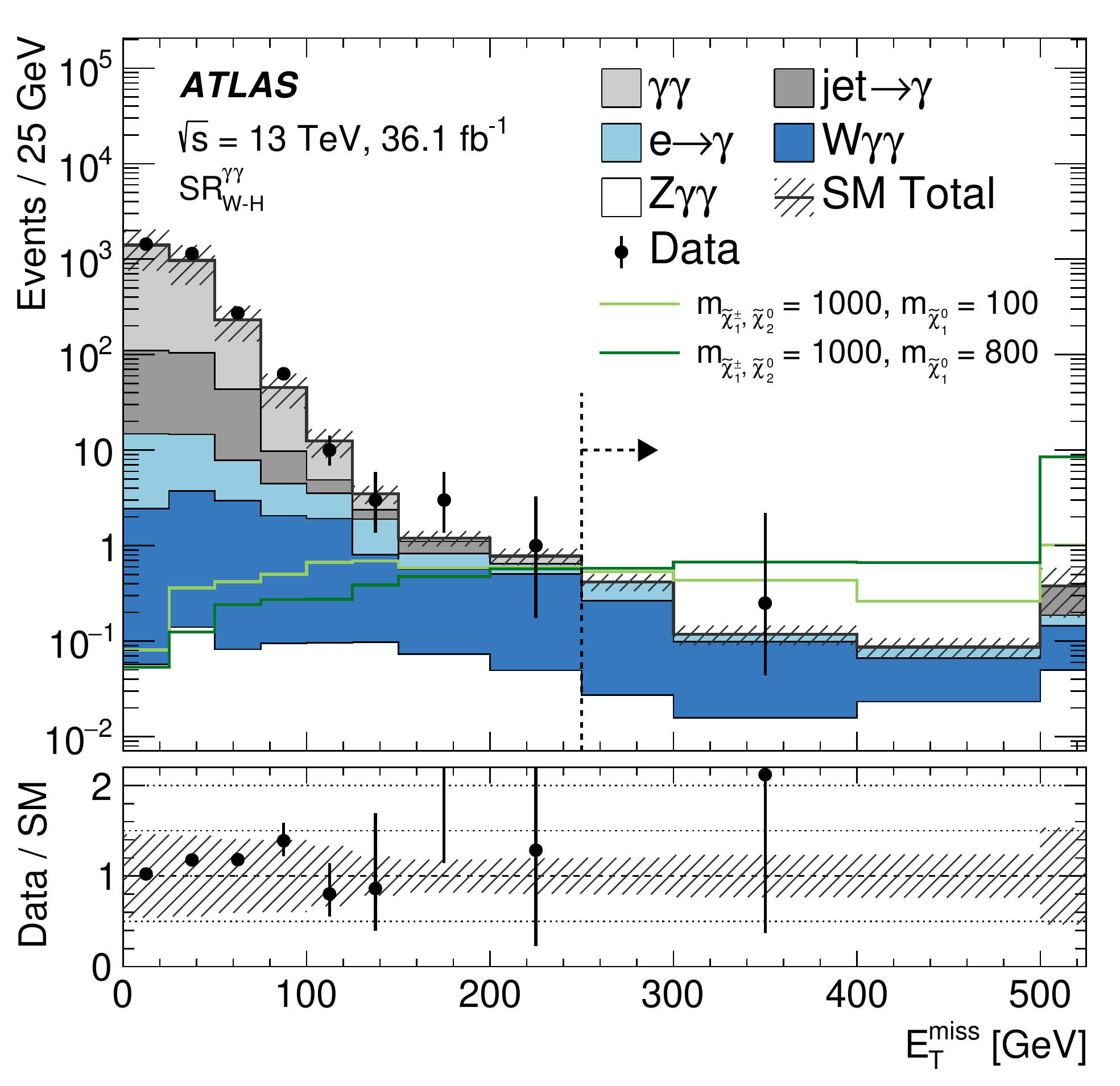}
  \end{center}
  \caption{Distribution of the missing transverse momentum \MET for the sample satisfying all
requirements of the (left) \BWL and (right) \BWH selection except the \MET requirement itself. Overlaid
are the expected SM backgrounds, separated into
the various contributing sources. Also shown are the signal expectations
for the $(m_{\wino},m_{\neutralino}) = (1000,100)$~GeV
and $(m_{\wino},m_{\neutralino}) = (1000,800)$~GeV models.
The vertical dashed lines and right-pointing arrows show the region of the
\MET observable selected for inclusion in \BWL and \BWH.
The lower panels show the ratio of observed data to the combined SM expectation.
For these plots, the band represents the range of combined statistical and systematic uncertainty in the SM expectation.
Events outside the range of the displayed region are included in the highest-value bin.
    \label{fig:diphoton_met}
  }
\end{figure}

Figure~\ref{fig:diphoton_met} shows the distribution of the missing transverse momentum \MET for the sample satisfying all
requirements of the \BWH (left) and \BWL (right) selections except the \MET requirement itself. Overlaid
are the expected SM backgrounds, separated into
the various contributing sources.

\begingroup
\renewcommand*{\arraystretch}{1.00}
\begin{table}[t]
  \caption{Definition, expected content and observed content of the seven validation regions used to confirm
the diphoton analysis background model. Here, $N_{\mathrm{lep}}$ is the number of required leptons of the stated
type, and $N_{\mathrm{exp}}$ and $N_{\mathrm{obs}}$ are the expected and observed numbers of events, respectively.
The remainder of the quantities are defined in the text. Events satisfying the selection requirements of any of the 
four diphoton signal regions are excluded from these validation regions.
The uncertainties in the numbers of expected events are
the combined statistical and systematic uncertainties. An ellipsis is entered when no such requirement is
made of the given validation region. 
}
  \begin{center}
\sisetup{retain-explicit-plus}
    \begin{tabular*}{\textwidth}{@{\extracolsep{\fill}}lrrrrrcr} \hline\hline\noalign{\smallskip}
         & \Etg [GeV] & \dphijm & $N_{\mathrm{lep}}$ & \HT [GeV] & \MET [GeV]   & $N_{\mathrm{exp}}$        & $N_{\mathrm{obs}}$   \\
      \noalign{\smallskip}\hline\noalign{\smallskip}
VR1$^{\gamma\gamma}$      & $> 75$     & $> 0.5$ &   ...       &   ...          &  $< 150$      & $43500 \pm 4400$  & $43918$     \\
VR2$^{\gamma\gamma}$      & $> 75$     & $> 0.5$ &   ...       & 1000--2500     &  $< 150$      & $ 2850 \pm  520$  & $ 3139$     \\
VR3$^{\gamma\gamma}$      & $> 75$     & $> 0.5$ &   ...       &   ...          &  100--150     & $  112 \pm   36$  & $  109$     \\
VR4$^{\gamma\gamma}$      & $> 50$     &   ...   &   $1e$      & $< 2000$       &    ...        & $ 34.5 \pm  7.2$  & $   38$     \\
VR5$^{\gamma\gamma}$      & $> 50$     &   ...   &   $1\mu$    & $< 2000$       &    ...        & $ 19.8 \pm  7.1$  & $   25$     \\
VR6$^{\gamma\gamma}$      & $> 75$     & $> 0.5$ &   ...       & $> 1750$       &    ...        & $  \phantom{1}290 \pm  130$  & $  336$     \\
VR7$^{\gamma\gamma}$      & $> 75$     & $> 0.5$ &   ...       &   ...          &  $> 100$      & $  139 \pm   40$  & $  146$     \\
   \hline\hline
    \end{tabular*}
    \label{tab:diphoton_vr}
  \end{center}
\end{table}
\endgroup

\subsection{Backgrounds to the photon+jets analysis}
\label{sec:sub_hbback}

Backgrounds from SM contributions to the three photon+jets SRs are expected to arise from both
events with real photons and events for which an electron or a
jet is misidentified as a photon. The former source is
expected to receive contributions from events in which a $W/Z$ boson
or a \ttbar pair is produced in association with a real photon ($W\gamma$, $Z\gamma$ and $\ttbar\gamma$ backgrounds), with
neutrinos in the subsequent weak decays of these produced states
providing significant \MET. The contribution from single-top production in association with a high-energy photon is
expected to be negligible.
Events with real photons can also contribute to the background in the photon+jets analysis when significant \MET
arises from instrumental sources (QCD background).
The $W\gamma$, $\ttbar\gamma$ and QCD backgrounds are estimated by constraining a
corresponding MC sample to match the observed event count in a dedicated CR
enriched in the given background process but otherwise kinematically
similar to the given SR, making use of the maximum-likelihood approach described at the beginning
of this section.
The MC simulation is then used
to provide an estimate of the expected background in the photon+jets
SRs. Smaller contributions from $Z$$\gamma$ and $\gamma\gamma$ (with or without an
accompanying $W$ or $Z$ boson) production
are estimated
directly from the MC simulation. The methods used to estimate
contributions from events for which electrons (``$e \rightarrow \gamma$'' backgrounds)
or jets (``jet $\rightarrow \gamma$'' backgrounds) are misidentified as photons
are identical to those used in the diphoton analysis, with the exception
that the single-photon trigger sample is used instead of the diphoton trigger sample, 
the requirement that the electron or loose photon be accompanied by a
tight isolated photon is removed, and the requirement for photons to be considered poorly isolated is changed to
$8 < \EtOFour - 0.22 \times \Etg - 2.45 < 27$~GeV.

All CRs require at least one isolated photon with $\et > 145$~GeV.
The QCD-background control region CR$_{\gamma+\text{jets}}$ is similar to \HBL, but with the
\MET requirement lowered to
$\MET > 100$~GeV, the \RT requirement removed, the number of required jets lowered to three,
and the \dphijm requirement inverted.
This provides a region dominated by real photons arising from radiative
QCD processes that is otherwise fairly similar to the photon+jets SRs.
The $W\gamma$-background control region CR$_{W\gamma}$
is defined by requiring that there be one or more
isolated leptons (electron or muon), at least one jet, and no $b$-tagged jet in the event. In addition, the \MET
requirement is changed to $100 < \MET < 200$~GeV and the \MEFF requirement reduced to $\MEFF > 500$~GeV
in order to enhance and isolate the $W\gamma$ contribution.
The $\ttbar\gamma$-background control region CR$_{t\bar{t}\gamma}$
is defined similarly, but requires at least two jets, and that two of the jets are $b$-tagged jets.
In order to increase the number of events in the CR the \MET requirement is lowered
to $50 < \MET < 200$~GeV.
Both the $W\gamma$-background and $\ttbar\gamma$-background CRs maintain the requirement
$\dphijm > 0.4$. Table~\ref{tab:phjet_cr} summarizes the selection criteria for the
three photon+jets analysis CRs.

\begin{table}[t]
  \centering

  \caption{Selection criteria for the three photon+jets analysis control regions. Here,
$N_{\gamma}$ is the number of required photons, $\et^{\gamma}$ the
transverse energy of the leading photon, $N_{\mathrm{lep}}$ the
number of required leptons, $N_{\mathrm{jets}}$ the number of required jets, 
and $N_{b}$$_{\text{-jets}}$ the number of required $b$-quark jets. 
The remainder of the quantities are defined in the text.
An ellipsis is entered when no such requirement is
made in the given control region.}

  \begin{tabular}{lrrr}
    \hline \hline
    &                          CR$_{\gamma+\text{jets}}$ & CR$_{W\gamma}$   & CR$_{t\bar{t}\gamma}$ \\
    \hline
    $N_{\gamma}$               &        $\ge1$ &          $\ge1$   &           $\ge1$ \\
    $\et^{\gamma}$      &    $>145$ $\gev$ &      $>145$ $\gev$   &       $>145$ $\gev$ \\
    $N_{\mathrm{lep}}$               &            0  &          $\ge1$   &           $\ge1$ \\
    $E_{\mathrm{T}}^{\mathrm{miss}}$     &  $>100$ $\gev$  &  100--200 $\gev$ &  50--200 $\gev$ \\
    $N_{\mathrm{jets}}$                  &        $\ge3$ &          $\ge1$   &           $\ge2$ \\
    $N_{b}$$_{\text{-jets}}$             &           ... &             $0$   &          $\ge 2$ \\
    $\Delta\phi(\text{jet}, \met)$       &        $<0.4$ &          $>0.4$   &           $>0.4$ \\
    $\Delta\phi(\gam, \met)$             &        $>0.4$ &             ...   &             ...  \\
    $m_{\mathrm{eff}}$                 &   $>2000$ $\gev$ &      $>500$ $\gev$   &       $>500$ $\gev$ \\
    \hline \hline
  \end{tabular}
    \label{tab:phjet_cr}
\end{table}

The event counts in the resulting QCD, $W\gamma$ and $\ttbar\gamma$ CRs are
used to scale the $\gamma$+jet, $W\gamma$ and $\ttbar\gamma$ MC samples, respectively, after applying a selection identical to
that of the corresponding CR.
The scale factors are determined in a simultaneous fit to all CRs,
taking into account mutual cross contamination between the different backgrounds.
The scale factors (ratio of the
derived background contribution in the corresponding control region to the MC expectation)
are found to be $1.67 \pm 0.49$,
$1.24 \pm 0.11$ and $1.20 \pm 0.17$ for the QCD, $W\gamma$ and $\ttbar\gamma$
backgrounds, respectively.
The resulting SR contributions from the QCD, $W\gamma$ and $\ttbar\gamma$ processes
depend upon transfer factors, given by MC
simulation, that relate the contribution of a given background process in the CR to that in the SR.
Uncertainties in the transfer factors
include those arising from experimental uncertainties in the efficiency for identifying
objects and in measuring their energy,
as well as theoretical uncertainties that are estimated by varying the underlying PDF set and renormalization
and factorization scales used in the generation of the MC background samples.
These uncertainties are incorporated into the overall background estimate uncertainties that arise from the simultaneous fit.
Estimates for the contributions of
the three real-photon backgrounds are shown in Table~\ref{tab:background_photon+j},
with the overall uncertainty taking into account correlations between the
various background sources.
For the three photon+jets SRs, the systematic uncertainty in each background estimate is
dominated by the theoretical uncertainties in the relevant MC samples
and the experimental uncertainties in the jet energy scale and resolution.

\begingroup
\renewcommand*{\arraystretch}{1.15}
\begin{table}[t]
  \caption{The expected and observed numbers of events in
    the photon+jets signal regions. The quoted errors are the combined
    statistical and systematic uncertainties.}

  \begin{center}
    \begin{tabular*}{\textwidth}{@{\extracolsep{\fill}}lccc} \hline\hline\noalign{\smallskip}
      {Signal region} & \HBL & \HBLP & \HBH  \\
      \noalign{\smallskip}\hline\noalign{\smallskip}
      $\gamma$ + jets (QCD)                          &   $0.00^{+0.21}_{-0.00}$ &   $0.42^{+0.43}_{-0.42}$ &          $0.14 \pm 0.14$ \\
      $W\gamma$                                      &          $0.54 \pm 0.24$ &          $0.81 \pm 0.22$ &          $0.40 \pm 0.26$ \\
      $Z\gamma$                                      &          $0.31 \pm 0.16$ &          $0.36 \pm 0.13$ &          $0.42 \pm 0.19$ \\
      $\ttbar \gamma$                                &          $0.30 \pm 0.11$ &          $0.54 \pm 0.17$ &          $0.07 \pm 0.03$ \\
      $e\rightarrow\gamma$                           &          $0.07 \pm 0.03$ &          $0.16 \pm 0.06$ &          $0.04 \pm 0.04$ \\
      $\mathrm{Jet}\rightarrow\gamma$                &   $0.07^{+0.44}_{-0.07}$ &   $0.35^{+0.36}_{-0.35}$ &   $0.01^{+0.50}_{-0.01}$ \\
      $\gamma\gamma/W\gamma\gamma/Z\gamma\gamma$     &          $0.03 \pm 0.01$ &          $0.03 \pm 0.01$ &          $0.06 \pm 0.02$ \\
      \hline
      Expected background events                     &   $1.33^{+0.58}_{-0.32}$ &   $2.68^{+0.64}_{-0.63}$ &   $1.14^{+0.61}_{-0.36}$ \\
      \hline
      Observed events & 4 & 8 & 3  \\
      \hline\hline
    \end{tabular*}
    \label{tab:background_photon+j}
  \end{center}
\end{table}
\endgroup

\begin{figure}[!htb]
  \begin{center}
    \includegraphics[width=0.96\textwidth]{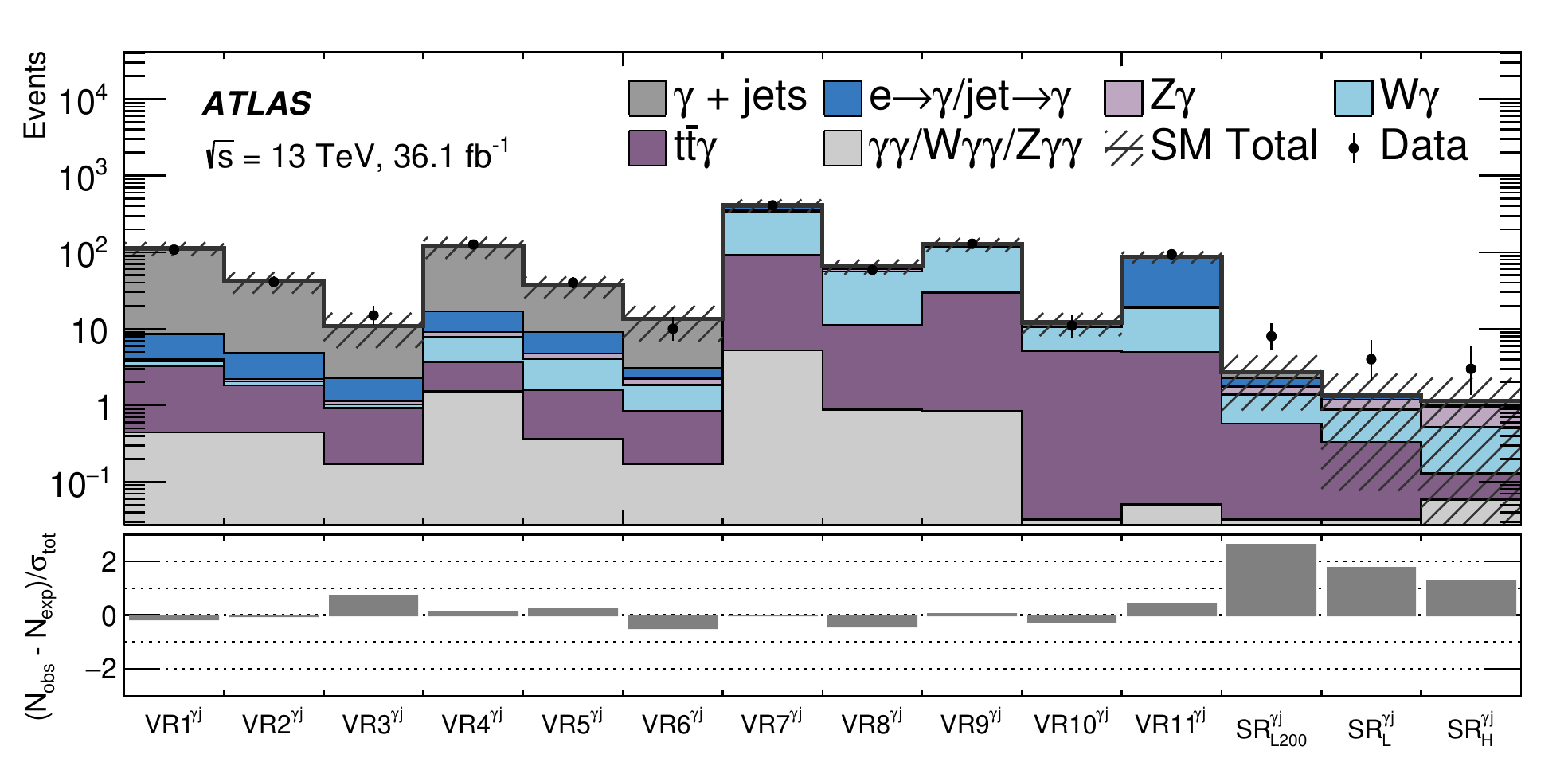}
  \end{center}
  \caption{Comparisons between expected and observed content of the
    validation and signal regions for the photon+jets analysis. 
The uncertainties in the expected numbers of events are
the combined statistical and systematic uncertainties.
The lower panel shows the pull (difference between observed and expected
event counts normalized by the uncertainty) for each region.
    \label{fig:photonjet_reg}
  }
\end{figure}

The accuracy of the resulting photon+jets analysis background model
is confirmed by the use of 11 VRs. Similar to the diphoton analysis VRs, these
VRs exclude events in
the various photon+jets SRs while having kinematic properties similar to those of the signal region.
Validation regions VR1$^{\gamma \mathrm{j}}$ through VR6$^{\gamma \mathrm{j}}$, defined in Table~\ref{tab:phjet_vr_gj}, target the
confirmation of the modeling of backgrounds arising from $\gamma$+jets production.
Validation regions VR7$^{\gamma \mathrm{j}}$ through VR11$^{\gamma \mathrm{j}}$, defined in Table~\ref{tab:phjet_vr_ew}, target the
confirmation of the modeling of backgrounds arising from $W \gamma$ and $\ttbar \gamma$ production
and from the misidentification of electrons as photons.
Figure~\ref{fig:photonjet_reg} shows the comparison between the expected and observed content
in the VRs,
with the expected content broken down into its contributing SM sources.

\begin{table}[t]
  \centering

  \caption{Definition, expected content, and observed content of the six validation regions used to confirm
the accuracy of the modeling of the $\gamma$ + jets background to the photon+jets analysis.
Here, $\et^{\gamma}$ is the
transverse energy of the leading photon, $N_{\mathrm{lep}}$ is the
number of required leptons, $N_{\mathrm{jets}}$ is the number of required jets,
and $N_{\mathrm{exp}}$ and $N_{\mathrm{obs}}$ are the expected and observed numbers of events, respectively.
The remainder of the quantities are defined in the text.
The uncertainties in the expected numbers of events are
the combined statistical and systematic uncertainties. An ellipsis is entered when no such requirement is
made in the given validation region.}

  \begin{tabular}{lrrrrrr}
  \hline \hline
                                       & VR1$^{\gamma \mathrm{j}}$ & VR2$^{\gamma \mathrm{j}}$ & VR3$^{\gamma \mathrm{j}}$ & VR4$^{\gamma \mathrm{j}}$ & VR5$^{\gamma \mathrm{j}}$ & VR6$^{\gamma \mathrm{j}}$ \\
  \hline
  $\et^{\gamma}$ [\gev] &            $> 145$ &           $> 145$ &          $> 145$ &            $> 400$ &          $> 400$ &       $> 400$ \\
  $N_{\mathrm{lep}}$               &                  0 &                 0 &                0 &                  0 &                0 &             0 \\
  $N_{\mathrm{jets}}$                  &             $\ge5$ &            $\ge5$ &           $\ge5$ &             $\ge3$ &           $\ge3$ &        $\ge3$ \\
  $\Delta\phi(\text{jet}, \met)$       &             $>0.4$ &            $>0.4$ &           $>0.4$ &             $>0.4$ &           $>0.4$ &        $>0.4$ \\
  $\Delta\phi(\gam, \met)$             &             $>0.4$ &            $>0.4$ &           $>0.4$ &             $>0.4$ &           $>0.4$ &        $>0.4$ \\
  ${\met}$ [\gev]                      &        50--175     &       75--175     &    100--175      &        100--175    &      125--175    &   150--175    \\
  ${\MEFF}$ [\gev]                     &           $> 2000$ &          $> 2000$ &         $> 2000$ &           $> 2000$ &         $> 2000$ &      $> 2000$ \\
  \RT                                  &            $<0.90$ &           $<0.90$ &          $<0.90$ &                ... &              ... &           ... \\
  \hline
  $N_{\mathrm{exp}}$                   & $112 \pm 20      $ & $42 \pm 11$       & $10.9 \pm 4.1$   & $120 \pm 36$       & $36.6 \pm 9.9$   & $13.4 \pm 5.5$ \\
  $N_{\mathrm{obs}}$                   &              $108$ &              $41$ &             $15$ &              $126$ &             $40$ &             $10$ \\
  \hline \hline
  \end{tabular}

  \label{tab:phjet_vr_gj}
\end{table}

\begin{table}[t]
  \centering

  \caption{Definition, expected content, and observed content of the five validation regions used to confirm
    the accuracy of the modeling of the $W \gamma$, $\ttbar \gamma$, and electron-to-photon misidentification
    backgrounds to the photon+jets analysis.
Here, $\et^{\gamma}$ is the
transverse energy of the leading photon, $N_{\mathrm{lep}}$ is the
number of required leptons, $N_{\mathrm{jets}}$ is the number of required jets, $N_{b}$$_{\text{-jets}}$
is the number of required $b$-quark jets, 
and $N_{\mathrm{exp}}$ and $N_{\mathrm{obs}}$ are the expected and observed numbers of events, respectively.
The remainder of the quantities are defined in the text.
    The uncertainties in the expected numbers of events are
    the combined statistical and systematic uncertainties. An ellipsis is entered when no such requirement is
    made in the given validation region.}

  \begin{tabular}{lrrrrr}
  \hline \hline
                                          &  VR7$^{\gamma \mathrm{j}}$  &  VR8$^{\gamma \mathrm{j}}$ &   VR9$^{\gamma \mathrm{j}}$ &  VR10$^{\gamma \mathrm{j}}$ &  VR11$^{\gamma \mathrm{j}}$ \\
  \hline
  $\et^{\gamma}$ [\gev]    &           $> 145$  &           $> 145$ &            $> 145$ &            $> 145$ &         $> 145$   \\
  $N_{\mathrm{lep}}$                  &             $\ge1$ &            $\ge1$ &             $\ge1$ &             $\ge1$ &              ...  \\
  $N_{\mathrm{jets}}$                     &             $\ge2$ &            $\ge2$ &             $\ge2$ &             $\ge2$ &            $\ge1$ \\
  $N_{b}$$_{\text{-jets}}$                &                ... &              ...  &                ... &                ... &            $\ge1$ \\
  $\Delta\phi(\text{jet}, \met)$          &             $>0.4$ &            $>0.4$ &             $>0.4$ &             $<0.4$ &            $>0.4$ \\
  $\Delta\phi(\gam, \met)$                &                ... &              ...  &                ... &                ... &            $<0.4$ \\
  $E_{\mathrm{T}}^{\mathrm{miss}}$ [\gev] &             $<200$ &            $<200$ &             $>200$ &             $>200$ &            $>200$ \\
  ${\MEFF}$  [\gev]                       &            $>1000$ &           $>1500$ &      $[1000,2000]$ &            $>1500$ &     $[500, 2000]$ \\
  \hline
  $N_{\mathrm{exp}}$                      & $408 \pm 79$       & $66    \pm 12$    & $127 \pm 23$       &   $12.1 \pm 2.1$   & $87 \pm 12$       \\
  $N_{\mathrm{obs}}$                      &              $410$ &              $59$ &              $129$ &               $11$ &              $94$ \\
  \hline \hline
  \end{tabular}
  \label{tab:phjet_vr_ew}
\end{table}

\begin{figure}[!hbt]
  \begin{center}
    \includegraphics[width=0.48\textwidth]{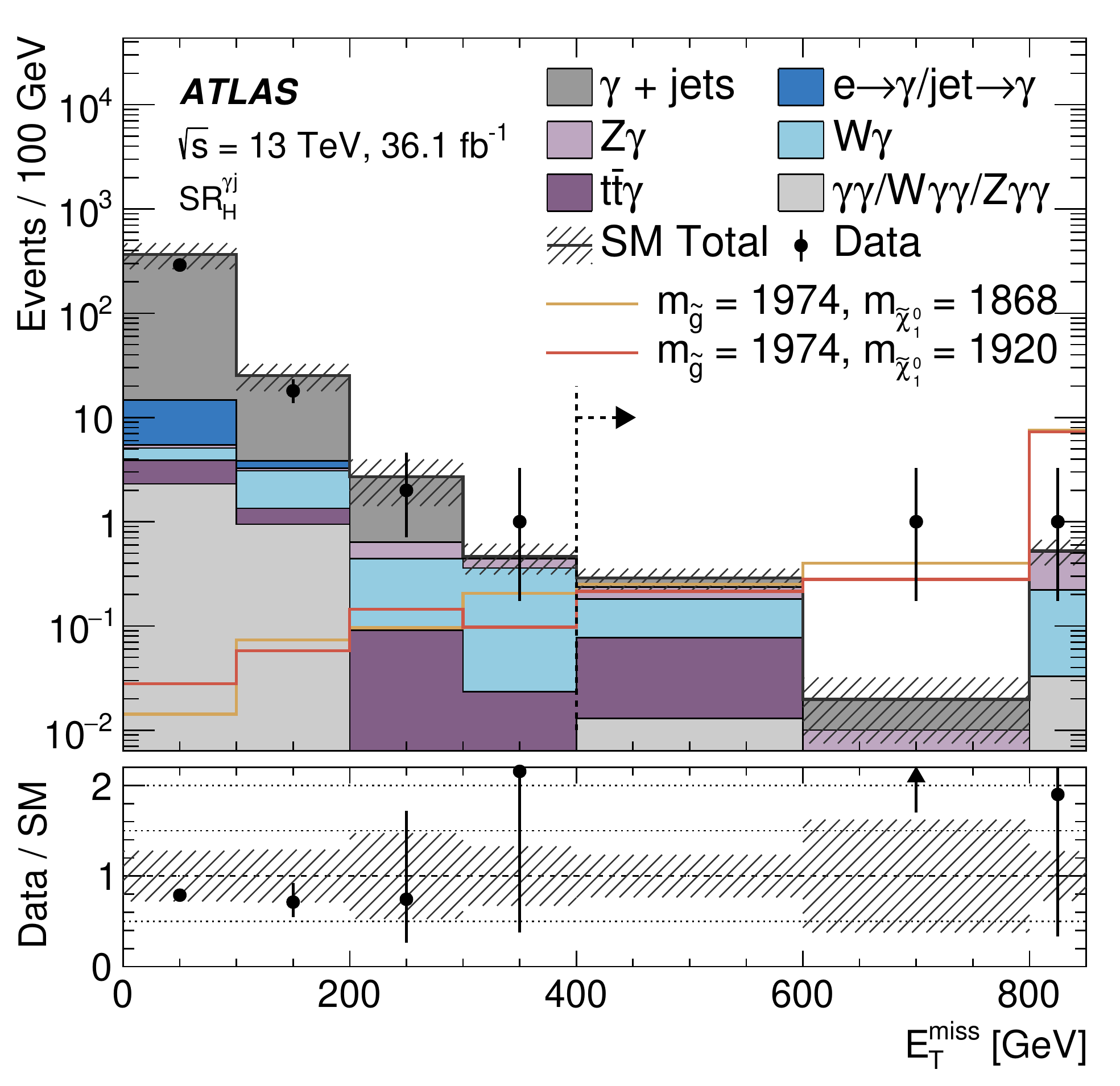} ~~
    \includegraphics[width=0.48\textwidth]{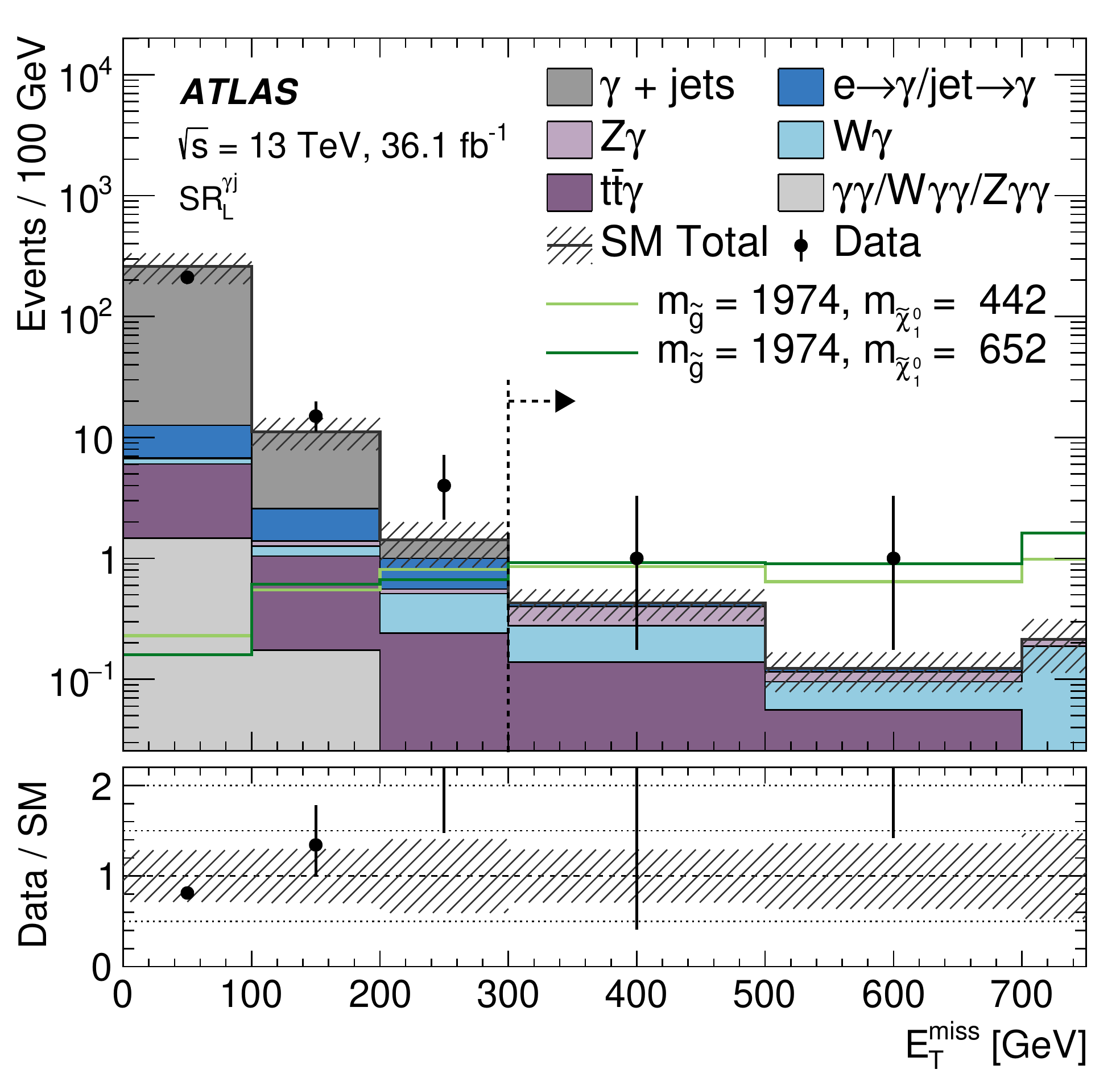}
  \end{center}
  \caption{
    Distribution of the missing transverse momentum \MET for the sample satisfying
    all requirements of the (left) \HBH and (right) \HBL or \HBLP selection except the \MET
    requirement itself. Overlaid
    are the expected SM backgrounds, separated into
    the various contributing sources. Also shown are the signal expectations
    for points in the $m_{\gluino}$--$m_{\neutralino}$ parameter space of the
    GGM model relevant to the photon+jets analysis (mass values in GeV).
The value of the gluino mass arises from the
    choice $M_3 = 1900$~GeV. The \neutralino mass values of 1868, 1920, 442 and 652~GeV arise from the
    choices $\mu =$ 1810, 1868, 400 and 600~GeV, respectively, combined with the constraint
    that the branching fraction of $\neutralino \rightarrow \gamma \gravitino$ be 50\%.
The vertical dashed lines and right-pointing arrows show the region of the
\MET observable selected for inclusion in \HBH and \HBL; for \HBLP, the \MET
requirement is \unit[200]{GeV} rather than \unit[300]{GeV}.
The lower panels show the ratio of observed data to the combined SM expectation.
For these plots, the band represents the range of statistical uncertainty in the SM expectation.
Events outside the range of the displayed region are included in the highest-value bin.
}
  \label{fig:photon_j_met}
\end{figure}

Figure~\ref{fig:photon_j_met} shows the distribution of the missing transverse momentum \MET for the sample satisfying
all requirements of the \HBH (left) and \HBL or \HBLP (right) selections except the \MET
requirement itself. Overlaid
are the expected SM backgrounds, separated into
the various contributing sources.

\section{Signal yield and associated uncertainties}
\label{sec:sig_eff}
GGM signal acceptances and efficiencies are estimated using MC simulation for each simulated
point in the gluino--bino, wino--bino, squark--bino and higgsino--bino parameter spaces,
and vary widely across the regions of these spaces relevant to establishing the model constraints
presented below. The product of acceptance and efficiency tends to be greatest ($30$\%--$35$\%)
when the masses of
both the produced and the NLSP states are largest, leading to large amounts of both visible energy and
missing transverse momentum that would clearly distinguish signal from background events. However,
for the more restrictive selection of the photon+jets analysis, particularly when
the NLSP mass is small, the product of
acceptance and efficiency can be significantly smaller.
For example, for the region relevant to establishing limits at low values of
$m_{\neutralino}$, the acceptance times efficiency of the \HBL selection is of the order of 0.1\%,
leading to a relatively modest constraint
on the mass of produced SUSY states.

The MC-based estimate of the signal yield is affected by 
various experimental systematic uncertainties, described below.
The resulting experimental systematic uncertainty in the signal
yield is incorporated in the determination of limits on the mass parameters
of the various GGM signal models considered in this search.

The uncertainty in the integrated luminosity is 2.1\%. It is derived, following a methodology similar 
to that detailed in Ref.~\cite{DAPR-2013-01}, from a calibration of the luminosity scale using $x$--$y$ beam-separation 
scans performed in August 2015 and May 2016.
Making use of a bootstrap method, the efficiency of the single-photon trigger is determined 
to be greater than 99\%, with an uncertainty of less than $\pm 1$\%,
for photons satisfying the photon+jets selection criteria~\cite{TRIG-2016-01}.
The diphoton trigger efficiency is found to be close to 100\% for events 
satisfying the diphoton analysis selection criteria, with an uncertainty of less than $\pm 0.4$\%.

The $\eta$-dependent uncertainty in the efficiency of photon identification, determined
as described in Ref.~\cite{PERF-2013-04}, is
between $\pm 0.2$\% and $\pm 0.4$\% for $\Etg < 200$~GeV, and between $\pm 1$\% and $\pm 4$\% for
larger values of \Etg.
The uncertainty in the energy scale for electrons and photons with high \ET, determined as described in Ref.~\cite{PERF-2013-05}, 
varies with $\eta$ over the range $\pm (0.5$--$1.5)$\%.
For high \ET, the uncertainty in the photon energy resolution is dominated by the uncertainty in
the constant term of the calorimetric energy resolution; at $\ET = 300$~GeV, the relative uncertainty is
$\pm (30$--$40)$\% depending on $\eta$.
For jets with $100 < \pt < 500$~GeV,
the uncertainty in the jet energy scale is found to be less
than $\pm 1$\%~\cite{PERF-2016-04}. Due to uncertainties in corrections for pileup, this uncertainty rises with falling \pt, reaching
a value of about $\pm 4.5$\% at $\pt = 20$~GeV.
Uncertainties in the values of whole-event observables, such as \MET and \HT, arise from
uncertainties in the energy of the objects from which they are constructed. In addition,
the \MET observable receives a contribution from tracks associated with the primary vertex but not associated with
any of the reconstructed objects in the event~\cite{PERF-2014-04}. Uncertainties arising from the inclusion of these unassigned
contributions are found to contribute negligibly to the overall uncertainty in the value of
the \MET observable.

In the regions of GGM parameter space relevant for establishing the exclusion
limits discussed in Section~\ref{sec:result}, and
excepting MC statistical uncertainty, the quadrature sum of the individual sources of systematic 
uncertainty in the signal reconstruction
efficiency in the diphoton analysis is
of order $\pm 5$\%, and is dominated by the uncertainties in photon identification and
the calorimetric energy scales. 
In the photon+jets analysis the systematic uncertainty is
larger (approximately $\pm 20$\%), due partially to an increased sensitivity to the jet energy scale
and resolution associated with the multiple-jet requirement.

\section{Results}
\label{sec:result}
\begingroup
\renewcommand*{\arraystretch}{1.15}
\begin{table}[t]
 \caption{Summary of the observed number of events
($N_\mathrm{obs}$), and the number of events expected from SM sources
($N_\mathrm{exp}$), for each of the seven SRs.
Also shown are the derived ($S_\mathrm{obs}^{95}$) and expected ($S_\mathrm{exp}^{95}$) model-independent 95\% C.L. limits on the number
of events from non-SM processes, and the observed ($\langle A \epsilon\mathrm{\sigma}\rangle_\mathrm{obs}^{95}$)
and expected ($\langle A \epsilon\mathrm{\sigma}\rangle_\mathrm{exp}^{95}$) 95\% C.L. limits on the visible
cross section from non-SM processes. The last column of the table shows 
the significance $Z$ of the observed excess (if any),
and the probability $p$, capped at 0.5, that an experiment with only
background fluctuates to at least the observed number of events.
   }
  \label{tab:mod_ind_lim}
\begin{center}
\setlength{\tabcolsep}{0.0pc}
\begin{tabular*}{\textwidth}{@{\extracolsep{\fill}}lccrcccl}
\noalign{\smallskip}\hline\hline\noalign{\smallskip}
Signal region                    & $N_\mathrm{obs}$ & $N_\mathrm{exp}$ & $S_\mathrm{obs}^{95}$ & $S_\mathrm{exp}^{95}$ & $\langle A \epsilon\mathrm{\sigma}\rangle_\mathrm{obs}^{95}$[fb] & $\langle A \epsilon\mathrm{\sigma}\rangle_\mathrm{exp}^{95}$[fb] & $Z$ ($p$)   \\
\noalign{\smallskip}\hline\noalign{\smallskip}
\BSL    &  0  & $0.50^{+0.30}_{-0.26}$ & 3.0  & $3.1 ^{+1.4}_{-0.2}$ & 0.083  & $0.086 ^{+0.039}_{-0.003}$   & 0.00 (0.50) \\
\BSH    &  0  & $0.48^{+0.30}_{-0.25}$ & 3.0  & $3.1 ^{+1.3}_{-0.1}$ & 0.083  &  $0.086 ^{+0.036}_{-0.003}$ & 0.00 (0.50) \\
\BWL    &  6  & $3.7 \pm 1.1$          & 8.6  & $5.8 ^{+2.8}_{-1.6}$  & 0.238  & $0.161 ^{+0.078}_{-0.044}$  & 1.06 (0.14) \\
\BWH    &  1  & $2.05^{+0.65}_{-0.63}$ & 3.7  & $4.4 ^{+1.9}_{-1.0}$ & 0.103  & $0.122 ^{+0.053}_{-0.028}$  & 0.00 (0.50) \\

\HBL    &  4  & $1.33^{+0.54}_{-0.32}$ &  7.6  & $4.7 ^{+1.6}_{-0.8}$ & 0.210  & $0.130 ^{+0.044}_{-0.022}$ &  1.81 (0.035) \\ 
\HBLP   &  8  & $2.68^{+0.64}_{-0.63}$ & 11.5  & $5.4 ^{+2.2}_{-1.2}$ & 0.318  & $0.151 ^{+0.060}_{-0.033}$ &  2.36 (0.009) \\ 
\HBH    &  3  & $1.14^{+0.61}_{-0.36}$ &  6.6  & $5.9 ^{+1.8}_{-1.1}$ & 0.183  & $0.162 ^{+0.050}_{-0.030}$ &  1.20 (0.116) \\ 

\noalign{\smallskip}\hline\hline\noalign{\smallskip}
\end{tabular*}
\end{center}
\end{table}
\endgroup

The number of events observed in each SR
is shown in Table~\ref{tab:mod_ind_lim},
along with the size of the expected SM
background. These results are also illustrated 
in Figures~\ref{fig:diphoton_reg} and~\ref{fig:photonjet_reg},
with the expected background broken down into its contributing SM sources.
No significant evidence of physics beyond the SM is observed in any of the SRs.

The most significant excess relative to the expected background is
observed in \HBLP of the photon+jets analysis. Considering
both statistical and systematic uncertainty, and assuming that all
observed events are from SM sources, an observation of eight or more
events over an expected background of $2.68^{+0.64}_{-0.63}$ events
represents an upward fluctuation with a probability of occurrence of approximately 0.9\%.

Based on the observed and expected numbers of events in the seven SRs shown in Table~\ref{tab:mod_ind_lim},
95\% C.L. upper limits are set for each SR on the number of events
from any scenario of physics beyond the SM. These limits are based on the
profile likelihood ratio~\cite{ref:histfit} and C.L.$_{\mathrm{s}}$~\cite{Read:2002hq} prescriptions, making
use of the likelihood function described in Section~\ref{sec:background}.
Assuming that no events due to physical processes beyond those
of the SM populate the various CRs used to estimate SR backgrounds,
observed 95\% C.L. upper limits on the number of such events vary between 3.0
(for \BSH and \BSL) and 11.5 (for \HBLP).
Dividing by the
integrated luminosity of \integLumi,
these number-of-event limits translate into 95\% C.L. upper
limits on the visible cross section for new physics, defined as the
product of cross section, branching fraction, acceptance and
efficiency, for the different SR definitions.
Here, the acceptance ($A$) is defined to be the fraction of
events whose underlying objects pass all kinematic and
whole-event selection requirements, and the efficiency ($\epsilon$)
to be the fraction of those events that would be observed
after reconstruction in the detector.
The resulting observed visible cross-section limits vary between
0.083~fb and 0.32~fb.

By considering, in addition to the event counts in the SRs, the values and uncertainties of the
acceptance times efficiency of the SR selection requirements,
as well as the NLO (+NLL) GGM
cross sections~\cite{Beenakker:1996ch,Kulesza:2008jb,Kulesza:2009kq,Beenakker:2009ha,Beenakker:2011fu,Fuks:2012qx,Fuks:2013vua},
95\% C.L. lower limits are set on the masses of the accessible SUSY states of 
the GGM scenarios explored in this study.
The SR with the
best expected sensitivity at each simulated point in the parameter space
of the corresponding GGM model(s) is used to determine the
degree of exclusion of that model point.

\begin{figure}[!ht]
  \begin{center}
    \includegraphics[width=0.80\textwidth]{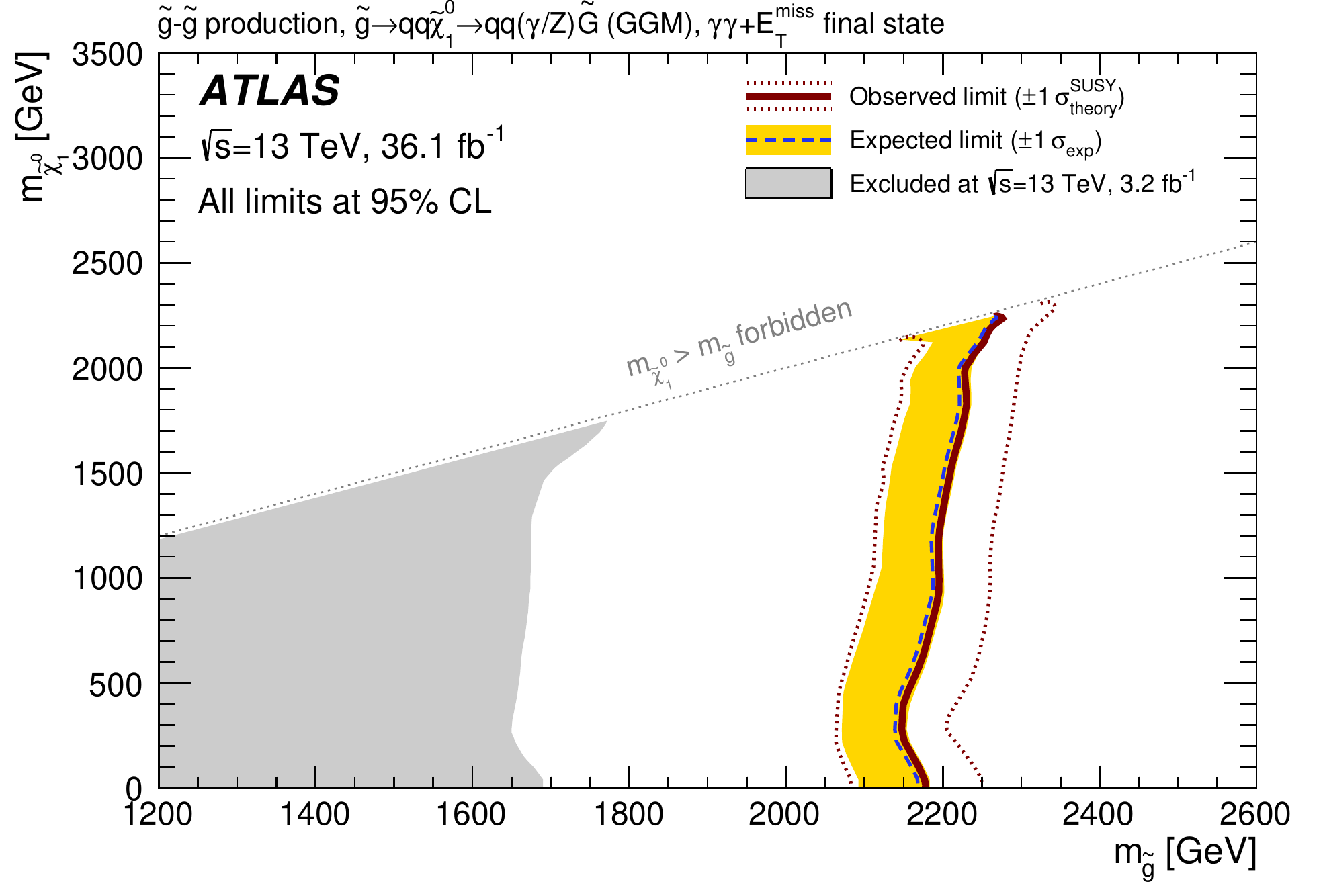}
\caption{
Exclusion limits in the gluino--bino mass plane, using the \BSH analysis for
$m_{\neutralino} > \GGMsplitG$~GeV and the \BSL analysis for
$m_{\neutralino} < \GGMsplitG$~GeV.
Combinations of gluino and bino mass are excluded at greater than 95\% C.L.
in the area to the left of the unbroken curve.
The observed limits are exhibited for the nominal SUSY model cross-section
expectation, as well as for a SUSY cross section increased and decreased
by 1 standard deviation of the cross-section systematic uncertainty.
Also shown is the expected limit, as well as the $\pm 1$ standard-deviation
range of the expected limit, which is asymmetric due to the small expected number of events.
The gray region is that previously excluded with the 2015 data sample; see Ref.~\cite{SUSY-2016-04}.
}
    \label{fig:di_gluino_limits}
  \end{center}
\end{figure}

\begin{figure}[!ht]
  \begin{center}
    \includegraphics[width=0.80\textwidth]{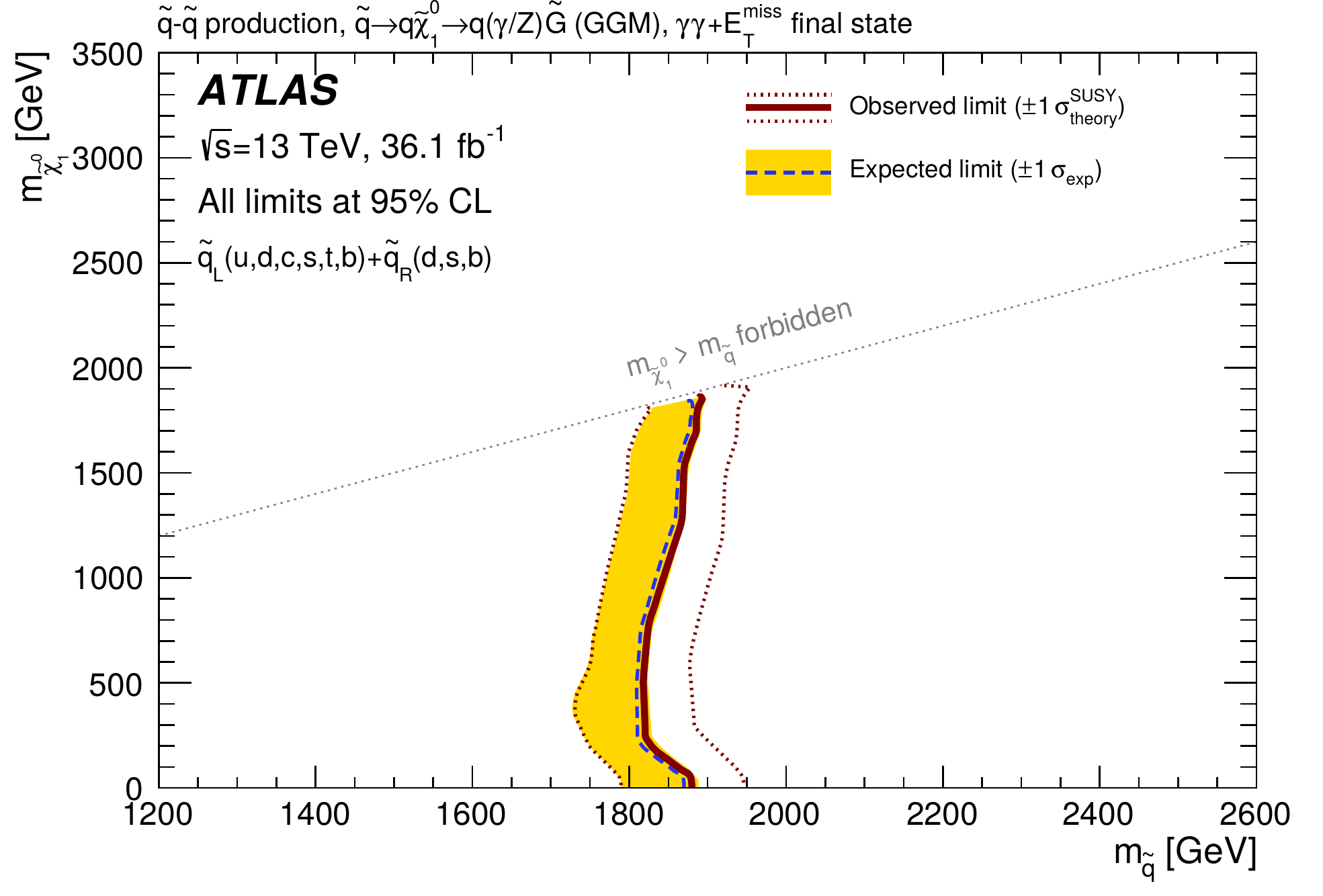}
    \caption{
Exclusion limits in the squark--bino mass plane, using the \BSH analysis for
$m_{\neutralino} > \GGMsplitS$~GeV and the \BSL analysis for
$m_{\neutralino} < \GGMsplitS$~GeV.
Combinations of squark and bino mass are excluded at greater than 95\% C.L.
in the area to the left of the unbroken curve.
The observed limits are exhibited for the nominal SUSY model cross-section
expectation, as well as for a SUSY cross section increased and decreased
by 1 standard deviation of the cross-section systematic uncertainty.
Also shown is the expected limit, as well as the $\pm 1$ standard-deviation
range of the expected limit, which is asymmetric due to the small number of expected events.
}
    \label{fig:di_squark_limits}
  \end{center}
\end{figure}

\begin{figure}[!ht]
  \begin{center}
    \includegraphics[width=0.80\textwidth]{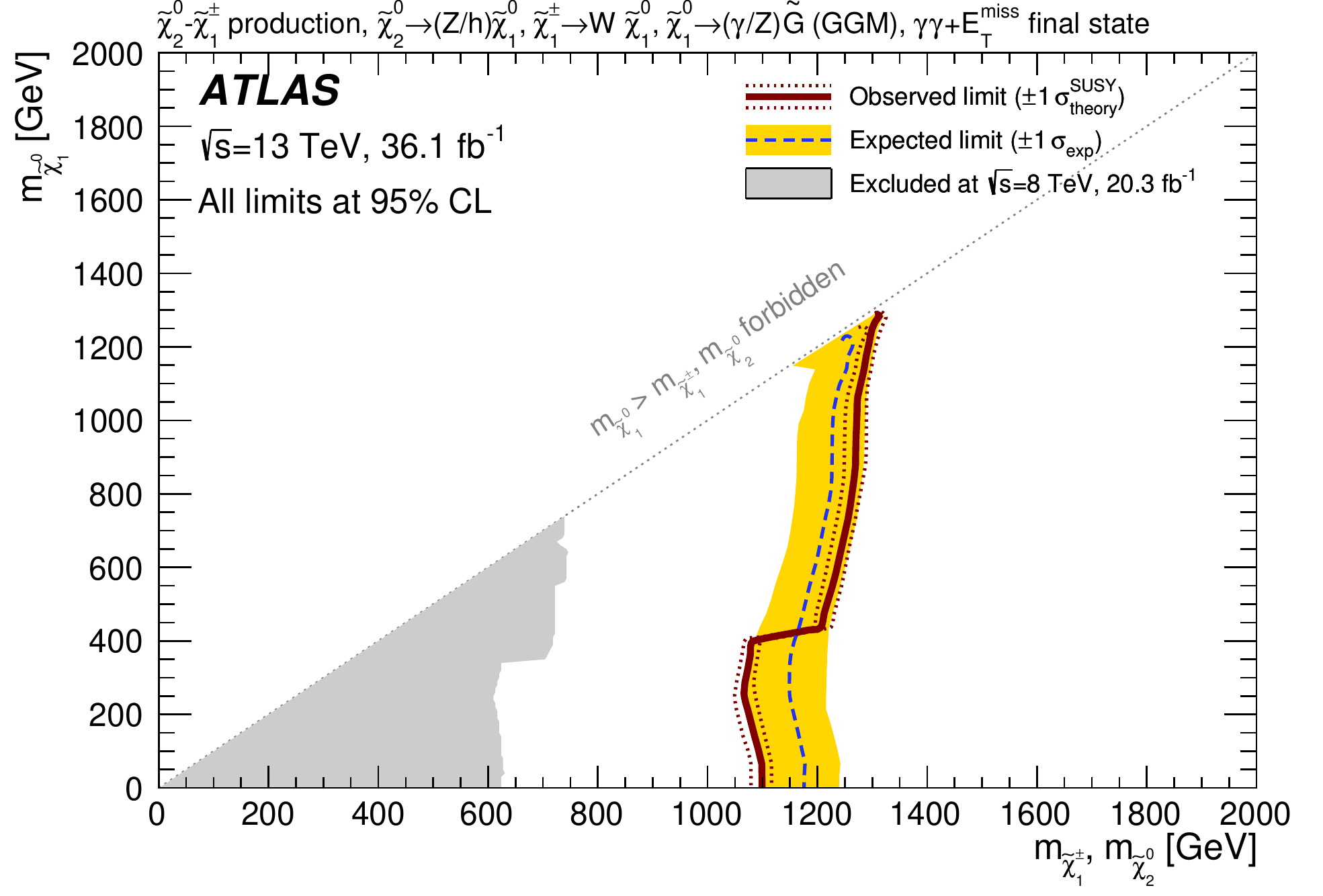}
    \caption{
Exclusion limits in the wino--bino
mass plane, using the \BWH analysis for
$m_{\neutralino} > \GGMsplitW$~GeV and the \BWL analysis for
$m_{\neutralino} < \GGMsplitW$~GeV.
The vertical axis represents bino mass while the horizontal axis represents wino mass.
Combinations of wino and bino masses are excluded at greater than 95\% C.L.
in the area to the left of the unbroken curve.
The observed limits are exhibited for the nominal SUSY model cross-section
expectation, as well as for a SUSY cross section increased and decreased
by 1 standard deviation of the cross-section systematic uncertainty.
Also shown is the expected limit, along with its $\pm 1$
standard-deviation range.
The discontinuity at $m_{\neutralino} = \GGMsplitW$~GeV is due to the
switch between the use of the \BWL and \BWH analyses, the former
of which exhibits a small excess of observed events relative to
the expected SM background.
The gray region is that previously excluded with the data sample
taken at $\sqrt{s} = 8$~TeV; see Ref.~\cite{SUSY-2014-01}.
}
    \label{fig:di_wino_limits}
  \end{center}
\end{figure}

For the diphoton analysis, in the region of gluino (squark) mass near the expected 
95\% C.L. exclusion limit, \BSH is expected to provide the
greatest sensitivity to the gluino--bino (squark--bino) model for bino masses
above \unit[\GGMsplitG]{GeV} (\unit[\GGMsplitS]{GeV}), with a transition to \BSL for bino masses below this value.
For the wino--bino model, the similar transition point between
the use of \BWL and \BWH is found to be at \unit[\GGMsplitW]{GeV}.
The resulting observed limits on the gluino and wino masses
are exhibited, as a function
of bino mass, for the diphoton analysis gluino, squark and wino production models in
Figures~\ref{fig:di_gluino_limits} through~\ref{fig:di_wino_limits}, respectively.
For the wino production model, the discontinuity at $m_{\neutralino} = \GGMsplitW$~GeV is due to the
small excess of events observed in the \BWL signal region.

\begin{figure}[!ht]
  \begin{center}
    \includegraphics[width=0.80\textwidth]{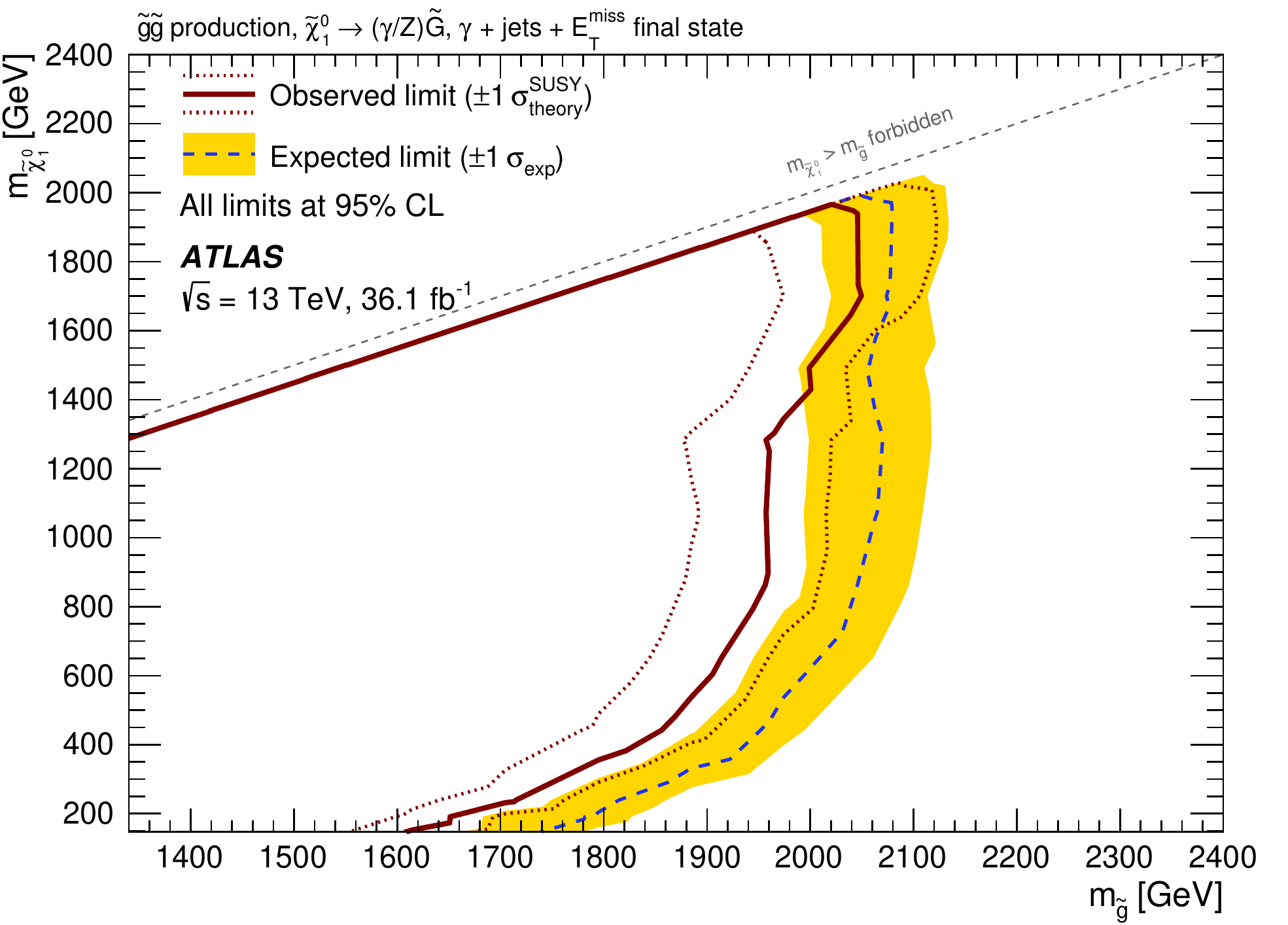}
    \caption{ Derived exclusion limits for the $\mu > 0$ higgsino--bino GGM model explored by the photon+jets analysis.
For this figure, the underlying model parameters
$M_3$ and $\mu$ have been transformed to the physical parameters
$m_{\gluino}$ and $m_{\neutralino}$, subject to the assumptions stated in Section~\ref{sec:susy}.
For each point in the higgsino--bino parameter space, the SR
(\HBL or \HBH) that provides the best expected sensitivity is used
to estimate the exclusion likelihood.
Combinations of gluino and neutralino mass are excluded at greater than 95\% C.L.
in the area to the left of the unbroken curve.
The observed limits are shown for the nominal SUSY model cross-section
expectation, as well as for a SUSY cross section increased and decreased
by 1 standard deviation of the cross-section systematic uncertainty.
The expected limit is also shown, along with its $\pm 1$ standard-deviation range.
}
    \label{fig:gamma_j_limits}
  \end{center}
\end{figure}

For the purpose of establishing these model-dependent limits, both the normalization of the \Wgg background estimate
and the limit on the possible number of events from new physics
are extracted from a simultaneous fit to the SR and \Wgg control region.
However, for masses near the various diphoton-analysis exclusion limits, the signal contamination in the \Wgg control
sample is appreciable only for the wino--bino parameter space, reaching
approximately 0.4 events (4\% of the 9.1 events in the \lgg CR attributed to the
\Wgg process) as the bino mass
approaches zero.
Also shown in these three figures, as well as in Figure~\ref{fig:gamma_j_limits},
are the expected limits, including their
statistical and background uncertainty ranges,
as well as observed limits for SUSY model cross sections $\pm 1$ standard deviation of theoretical
uncertainty from their central value.

Considering all possible values of the \neutralino mass,
95\% C.L. lower limits of
\GGMlimitG, \GGMlimitS and \GGMlimitW
are set by the diphoton analysis on the value of the gluino, squark or wino mass, respectively,
for any value of the NLSP bino mass less than that
of the gluino, squark or wino mass.
Based on a sample of 35.9 fb$^{-1}$ of $pp$ data accumulated at
$\sqrt{s} = 13$~TeV, and assuming a branching fraction of 100\% for
the photonic decay of the \neutralino, the CMS Collaboration
has set 95\% C.L. lower limits of \unit[1790]{GeV} and \unit[1580]{GeV}
for similar models of gluino and squark production and decay, respectively~\cite{CMS-SUS-16-047}.
For a GGM model similar to the wino--bino model of the diphoton analysis, 
a separate CMS Collaboration analysis~\cite{CMS-SUS-16-047} has set a 95\% C.L. lower limit
as high as \unit[1000]{GeV} on the wino mass, depending on the value of the binolike \neutralino mass.

Using the photon+jets analysis, limits are set in the two-dimensional plane of the
masses of the gluino and the mixed higgsino--bino NLSP.
For values of $m_{\gluino}$ and $m_{\neutralino}$ close to the expected 95\% C.L. exclusion limit,
\HBL is
expected to provide a greater sensitivity for
NLSP masses below approximately \unit[1500]{GeV}, and so is made use of in this region;
for higher NLSP masses, \HBH is used to establish the degree of exclusion of points
in the GGM-model parameter space.
The resulting observed exclusion contour is shown in Figure~\ref{fig:gamma_j_limits}.
In the context of this GGM model, lower
limits as high as 
\unit[2050]{GeV} are established
for the gluino mass, depending on the value of $m_{\neutralino}$.
The sensitivity of the analysis has not been explored for values of the NLSP mass within
\unit[50]{GeV} of that of the gluino, where the selection efficiency diminishes due to the restriction
of phase space for producing multiple high-\pt jets, and the tendency of the
gluino to become metastable as the splitting between the gluino and \neutralino masses becomes small.

\FloatBarrier

\section{Conclusion}
\label{sec:conclusion}

Making use of proton--proton collision data at $\sqrt{s} = 13$ TeV corresponding to an integrated luminosity of \integLumi recorded by the ATLAS
detector at the LHC in 2015 and 2016, a search is performed for photonic
signatures of new physics associated with significant missing transverse momentum.
Single-photon and diphoton selection strategies were developed and used to
search for evidence for several general gauge-mediated SUSY-breaking scenarios.
No significant excess of events over the Standard Model expectation
is observed in any of the searches, and limits are
set on possible contributions of new physics. Model-independent limits 
between 0.083~fb and 0.32~fb are set
on the associated visible
cross section of contributions from physics beyond the Standard Model.

Based on these limits on contributions from new physics, 
model-dependent limits are set 
on the masses of SUSY particles
within the context of GGM.
A diphoton signature is used to search for strongly and weakly produced SUSY states with
a decay chain proceeding through a binolike next-to-lightest supersymmetric particle (NLSP). In the context of these
models, lower limits of
\GGMlimitG, \GGMlimitS and \GGMlimitW are set on the masses of gluinos,
squarks and
a degenerate set of winos, respectively, for any value of the bino mass less
than the mass of these produced states. 
In addition, a photon+jets signature is
used to search for an alternative scenario in which the GGM NLSP is a higgsino--bino
admixture with a roughly equal branching fraction to photons and
$Z$ bosons. 
In the context of this model, lower
limits as high as 
\unit[2050]{GeV} are established
for the gluino mass, depending on the value of the NLSP mass.


\section*{Acknowledgements}


We thank CERN for the very successful operation of the LHC, as well as the
support staff from our institutions without whom ATLAS could not be
operated efficiently.

We acknowledge the support of ANPCyT, Argentina; YerPhI, Armenia; ARC, Australia; BMWFW and FWF, Austria; ANAS, Azerbaijan; SSTC, Belarus; CNPq and FAPESP, Brazil; NSERC, NRC and CFI, Canada; CERN; CONICYT, Chile; CAS, MOST and NSFC, China; COLCIENCIAS, Colombia; MSMT CR, MPO CR and VSC CR, Czech Republic; DNRF and DNSRC, Denmark; IN2P3-CNRS, CEA-DRF/IRFU, France; SRNSFG, Georgia; BMBF, HGF, and MPG, Germany; GSRT, Greece; RGC, Hong Kong SAR, China; ISF, I-CORE and Benoziyo Center, Israel; INFN, Italy; MEXT and JSPS, Japan; CNRST, Morocco; NWO, Netherlands; RCN, Norway; MNiSW and NCN, Poland; FCT, Portugal; MNE/IFA, Romania; MES of Russia and NRC KI, Russian Federation; JINR; MESTD, Serbia; MSSR, Slovakia; ARRS and MIZ\v{S}, Slovenia; DST/NRF, South Africa; MINECO, Spain; SRC and Wallenberg Foundation, Sweden; SERI, SNSF and Cantons of Bern and Geneva, Switzerland; MOST, Taiwan; TAEK, Turkey; STFC, United Kingdom; DOE and NSF, United States of America. In addition, individual groups and members have received support from BCKDF, the Canada Council, CANARIE, CRC, Compute Canada, FQRNT, and the Ontario Innovation Trust, Canada; EPLANET, ERC, ERDF, FP7, Horizon 2020 and Marie Sk{\l}odowska-Curie Actions, European Union; Investissements d'Avenir Labex and Idex, ANR, R{\'e}gion Auvergne and Fondation Partager le Savoir, France; DFG and AvH Foundation, Germany; Herakleitos, Thales and Aristeia programmes co-financed by EU-ESF and the Greek NSRF; BSF, GIF and Minerva, Israel; BRF, Norway; CERCA Programme Generalitat de Catalunya, Generalitat Valenciana, Spain; the Royal Society and Leverhulme Trust, United Kingdom.

The crucial computing support from all WLCG partners is acknowledged gratefully, in particular from CERN, the ATLAS Tier-1 facilities at TRIUMF (Canada), NDGF (Denmark, Norway, Sweden), CC-IN2P3 (France), KIT/GridKA (Germany), INFN-CNAF (Italy), NL-T1 (Netherlands), PIC (Spain), ASGC (Taiwan), RAL (UK) and BNL (USA), the Tier-2 facilities worldwide and large non-WLCG resource providers. Major contributors of computing resources are listed in Ref.~\cite{ATL-GEN-PUB-2016-002}.

\clearpage


\printbibliography



\clearpage 
\begin{flushleft}
{\Large The ATLAS Collaboration}

\bigskip

M.~Aaboud$^\textrm{\scriptsize 137d}$,
G.~Aad$^\textrm{\scriptsize 88}$,
B.~Abbott$^\textrm{\scriptsize 115}$,
O.~Abdinov$^\textrm{\scriptsize 12}$$^{,*}$,
B.~Abeloos$^\textrm{\scriptsize 119}$,
S.H.~Abidi$^\textrm{\scriptsize 161}$,
O.S.~AbouZeid$^\textrm{\scriptsize 139}$,
N.L.~Abraham$^\textrm{\scriptsize 151}$,
H.~Abramowicz$^\textrm{\scriptsize 155}$,
H.~Abreu$^\textrm{\scriptsize 154}$
% ...
\bigskip
\\
% ...
$^{12}$ Institute of Physics, Azerbaijan Academy of Sciences, Baku, Azerbaijan\\
% ...
$^{88}$ CPPM, Aix-Marseille Universit{\'e} and CNRS/IN2P3, Marseille, France\\
% ...
$^{115}$ Homer L. Dodge Department of Physics and Astronomy, University of Oklahoma, Norman OK, United States of America\\
% ...
$^{119}$ LAL, Univ. Paris-Sud, CNRS/IN2P3, Universit{\'e} Paris-Saclay, Orsay, France\\
% ...
$^{137}$ $^{(a)}$ Facult{\'e} des Sciences Ain Chock, R{\'e}seau Universitaire de Physique des Hautes Energies - Universit{\'e} Hassan II, Casablanca; $^{(b)}$ Centre National de l'Energie des Sciences Techniques Nucleaires, Rabat; $^{(c)}$ Facult{\'e} des Sciences Semlalia, Universit{\'e} Cadi Ayyad, LPHEA-Marrakech; $^{(d)}$ Facult{\'e} des Sciences, Universit{\'e} Mohamed Premier and LPTPM, Oujda; $^{(e)}$ Facult{\'e} des sciences, Universit{\'e} Mohammed V, Rabat, Morocco\\
% ...
$^{139}$ Santa Cruz Institute for Particle Physics, University of California Santa Cruz, Santa Cruz CA, United States of America\\
% ...
$^{151}$ Department of Physics and Astronomy, University of Sussex, Brighton, United Kingdom\\
% ...
$^{154}$ Department of Physics, Technion: Israel Institute of Technology, Haifa, Israel\\
$^{155}$ Raymond and Beverly Sackler School of Physics and Astronomy, Tel Aviv University, Tel Aviv, Israel\\
% ...
$^{161}$ Department of Physics, University of Toronto, Toronto ON, Canada\\
% ...
$^{a}$ Also at Department of Physics, King's College London, London, United Kingdom\\
$^{b}$ Also at Institute of Physics, Azerbaijan Academy of Sciences, Baku, Azerbaijan\\
$^{c}$ Also at Novosibirsk State University, Novosibirsk, Russia\\
$^{d}$ Also at TRIUMF, Vancouver BC, Canada\\
$^{e}$ Also at Department of Physics {\&} Astronomy, University of Louisville, Louisville, KY, United States of America\\
% ...
$^{*}$ Deceased
\end{flushleft}


\end{document}